\documentclass{article}

\PassOptionsToPackage{numbers, compress}{natbib}
\usepackage[preprint]{neurips_2024}

\usepackage[utf8]{inputenc} 
\usepackage[T1]{fontenc}    
\usepackage{hyperref}       
\usepackage{url}            
\usepackage{booktabs}       
\usepackage{amsfonts}       
\usepackage{amsmath}
\usepackage{amsthm}
\usepackage{amssymb}
\usepackage{subfigure}
\usepackage{nicefrac}       
\usepackage{microtype}      
\usepackage{xcolor}         
\usepackage{url} 
\usepackage{bbm} 
\usepackage{color}
\usepackage{rotating}
\usepackage{makecell}
\usepackage{multirow}
\usepackage{float}
\usepackage[linesnumbered,ruled]{algorithm2e}
\usepackage{subfigure}
\usepackage{adjustbox}
\usepackage{wrapfig}
\newcommand{\bX}{{\bf X}}
\newcommand{\bx}{{\bf x}}

\newcommand{\E}{\mathbb{E}}

\newcommand{\bdelta}{\mbox{\boldmath{$\delta$}}}

\newcommand{\bbeta}{\mbox{\boldmath{$\beta$}}}

\newcommand{\bmu}{\mbox{\boldmath{$\mu$}}}

\newcommand{\bc}{\begin{center}}
\newcommand{\ec}{\end{center}}
\newcommand{\be}{\begin{equation}}
\newcommand{\ee}{\end{equation}}
\newcommand{\ba}{\begin{array}}
\newcommand{\ea}{\end{array}}
\newcommand{\bean}{\begin{eqnarray*}}
\newcommand{\eean}{\end{eqnarray*}}
\newcommand{\bea}{\begin{eqnarray}}
\newcommand{\eea}{\end{eqnarray}}

\newtheorem{lemma}{\bf Lemma}
\newtheorem{theorem}{\bf Theorem}

\newtheorem{proposition}{\bf Proposition}
\newtheorem{definition}{\bf Definition}
\newtheorem{remark}{\bf Remark}
\newcommand{\ben}{\begin{enumerate}}
\newcommand{\een}{\end{enumerate}}
\newcommand{\bed}{\begin{itemize}}
\newcommand{\eed}{\end{itemize}}

\title{Uniform Pessimistic Risk and its Optimal Portfolio}

\author{
  Sungchul~Hong\\
  Department of Statistics\\
  University of Seoul\\
  \texttt{shong@uos.ac.kr} 
  \And
  Jong-June Jeon\thanks{Corresponding author.}\\
  Department of Statistics\\
  University of Seoul\\
  \texttt{jj.jeon@uos.ac.kr} 
}

\begin{document}

\maketitle

\begin{abstract}
    The optimal allocation of assets has been widely discussed with the theoretical analysis of risk measures, and pessimism is one of the most attractive approaches beyond the conventional optimal portfolio model. The $\alpha$-risk plays a crucial role in deriving a broad class of pessimistic optimal portfolios. However, estimating an optimal portfolio assessed by a pessimistic risk is still challenging due to the absence of a computationally tractable model. In this study, we propose an integral of $\alpha$-risk called the \textit{uniform pessimistic risk} and the computational algorithm to obtain an optimal portfolio based on the risk. 
    Further, we investigate the theoretical properties of the proposed risk in view of three different approaches: multiple quantile regression, the proper scoring rule, and distributionally robust optimization. Real data analysis of three stock datasets (S\&P500, CSI500, KOSPI200) demonstrates the usefulness of the proposed risk and portfolio model.
\end{abstract}

\section{Introduction}\label{sec: intro}
The concept of risk measures determining the allocation of optimal assets in modern portfolio theory has continually evolved. The $\sigma$-risk \cite{markowitz1952portfolio}, historically the most extensively researched risk measure, is based on the uncertainty of future asset returns. This simple yet useful concept has led to numerous derivative studies on risk diversification and practical outcomes. However, concerns and skepticism regarding the effectiveness of the $\sigma$-risk have persistently arisen in actual financial markets.

Asymmetric responses of market participants to risk have been criticized as the most significant limitation of the Markowitz model \cite{bradley2003financial,Nock2011OnTP}. Furthermore, the Markowitz optimum is not well-defined for assets with heavy-tailed distributions of returns.  For these reasons, researchers have been developing new risk measures to overcome the limitations of the $\sigma$-risk. This effort has given rise to a new research direction called ``pessimistic risk'', which aims to explain how market participants, in their asymmetric risk-averse behavior, tend to avoid extreme risks.

One of the most popular pessimistic risks is the $\alpha$-risk, which is modulated by a parameter $\alpha$ to judge whether an event is favorable or not \cite{bassett2004pessimistic}. Thus, the optimality induced by the $\alpha$-risk depends on the selection of $\alpha$ \cite{taniai2012statistically}. Additionally, \cite{bassett2004pessimistic, belles2014beyond} considered a weighted average of $\alpha$-risks with only discrete $\alpha$s, and the portfolio based on it outperforms ones based on the single $\alpha$.
However, what values of the combination of $\alpha$s produce a realistic model accounting for the behavioral bias during more unfavorable events is unknown. 
\cite{acerbi2002spectral} proposed a general case of $\alpha$-risk with the notion of spectral risk measure by defining it as an integral of $\alpha$-risk on $\alpha \in (0, 1)$. Their idea established a family of pessimistic risks and expanded the range of applications. 
However, developing such an integral $\alpha$-risk model has been challenging because of the computational cost and the complexity of quantile modeling. Furthermore, while several synthesized risks incorporating pessimistic risks have been proposed, they are primarily limited to risk measurement with empirical quantile function and have not been directly applied in portfolio optimization \cite{wirch1999synthesis}.

Motivated by this practical necessity, we propose the uniform pessimistic risk (UPR) and its optimization algorithm for portfolio construction. UPR is an integrated risk equipped with the uniform measure on $(0, 1)$. First, we investigate the theoretical properties of UPR as a limit of weighted composite quantile risk and a special case of a proper scoring rule \citep{matheson1976scoring}. Next, as an application of UPR, we present the optimal portfolio minimizing UPR. Since minimizing UPR requires not only an optimal weight of assets but also the distribution of the optimal portfolio, we tackle the problem of estimating the distribution function by introducing a spline model. Section \ref{sec: proposal} presents that minimizing UPR leads to the estimation of a truthful cumulative distribution function (CDF) as an inverse of the quantile function. The uncertainty of the return attained from the estimated portfolio can be quantified from the estimated CDF.

Finally, we discuss whether the uniform probability measure used in UPR is sufficiently large to cover a general class of pessimistic risk. \cite{rahimian2019distributionally} elicited the relationship between the law invariant coherent risk measures and distributionally robust optimization (DRO). As UPR is a law invariant coherent risk measure \citep{kusuoka2001law, shapiro2013kusuoka}, its portfolio optimization is equivalent to DRO. There exists a proper set of probability measures for $\alpha \in (0, 1)$ such that UPR is the maximum risk measure. Thus, the UPR portfolio model is robust to distribution-shift under some probability measure set. 

The remainder of the paper is organized as follows: Section \ref{sec: review} reviews the pessimistic portfolio theory and related statistical models. Section \ref{sec: proposal} defines UPR and proposes an algorithm to estimate the optimal portfolio assessed by the proposed risk measure. 
Section \ref{sec: portfolio} proposes an algorithm to estimate the optimal portfolio where UPR can be minimized. Section \ref{sec: num} shows our method's performance by the data analyses conducted with synthetic and real data. 
Conclusion and limitations of this study follow in Section \ref{sec: conc}. All proofs, source codes, and datasets are provided in Supplementary Materials.
\section{Background} \label{sec: review}
\subsection{$\alpha$-risk and its Optimal Portfolio}
The $\alpha$-risk is a popular alternative measure to the $\sigma$-risk. The $\alpha$-risk can be 
applied to the analysis of a broader class of uncertain assets because it can be defined without the existence of the second moment. 
Let $Y$ be a continuous random variable representing 
the return of a portfolio and assume $\E|Y| <\infty$.
Denote its CDF and quantile function by $F$ and $G$. 
The $\alpha$-risk of $Y$ is defined by 
\bea \label{def: alpha_risk}
\varrho_{\alpha}(Y) = - \int_0^{1} G(t)d\nu_{\alpha}(t) = -\alpha^{-1}\int_0^{\alpha} G(t)dt.
\eea
where $\nu_{\alpha}(\cdot)$ is the CDF of the uniform distribution on $(0,\alpha)$ for $\alpha \in (0,1)$.\\ 
Since the weight of $G(t)$ depends on  $\nu_{\alpha}(t)$, $\nu_{\alpha}(t)$ is called the distortion function \cite{wang2000class}. For a continuous random variable, the $\alpha$-risk is also referred to as the expected shortfall \citep{acerbi2002expected}, the conditional value at risk (CVaR) \citep{pflug2000some, rockafellar2002conditional}, the tail conditional expectation \citep{artzner1999coherent}. Proposition \ref{prop: qr} shows the relation between the $\alpha$-risk and the $\alpha$-quantile risk.
\begin{proposition}[\citep{bassett2004pessimistic}] \label{prop: qr}
Let $\mu_0 = \E (Y)$ and $\ell_\alpha(x) = (\alpha - I(x<0))x$. Then,
\bean
\min_{\beta_0\in \mathbb{R}} \E(\ell_\alpha (Y - \beta_0)) = \alpha(\mu_0 + \varrho_{\alpha}(Y)).
\eean
\end{proposition}

Proposition \ref{prop: qr} implies that the optimal portfolio minimizing $\alpha$-risk can be obtained from fitting a quantile regression model. 
Let $\bX\in \mathbb{R}^p$ be a random vector denoting the returns of $p$ assets, and let $Y = \bX^\top \bbeta$ be the return of the portfolio with weight $\bbeta \in \mathbb{R}^p$. Denote the set of portfolios with the expected return $\mu_0$ by  $C = \{\bbeta\in \mathbb{R}^p: \E(\bX^\top \bbeta) = \mu_0, {\bf 1}^\top \bbeta = 1 \}$. Through Proposition 1, the optimization problem to minimize $\alpha$-risk equivalently is written by
\bea
\underset{\beta \in C}{\mbox{argmin}}
 \Big(\min_{\beta_0 \in \mathbb{R}} \E(\ell_\alpha (\bX^\top \bbeta - \beta_0)) \Big)= \underset{\beta \in C}{\mbox{argmin }} \varrho_{\alpha} (\bX^\top \bbeta). \label{eq: qr_model}
\eea
That is, the $\alpha$-risk and the quantile risk are equivalent to obtaining the optimal portfolio for a given expected returns.

\subsection{Pessimistic Risk and its Optimal Portfolio}
The pessimism in risk analysis can be understood as a regime change of the conventional optimality in portfolio management. Building on the concept of $\alpha$-risk, the pessimistic risk is introduced to more flexibly assign weights to the less favorable events.  
\begin{definition}[\cite{bassett2004pessimistic}] \label{def: prm}
A risk measure $\varrho$ is pessimistic if, for a CDF $\varphi$ on $(0,1)$,
\bea 
\varrho(Y; \varphi) = \int^1_0 \varrho_{\alpha}(Y)d\varphi(\alpha).
\eea
\end{definition}
For a discrete measure $\varphi_d$ on $\{\alpha_k \in (0,1): \alpha_1< \cdots< \alpha_K\}$, where $\varphi_d(\{\alpha_k\}) = w_k$ for $w_k>0$ and $\sum_{k=1}^K w_k = 1$, the pessimistic risk is given by $\varrho (Y; \varphi_d) = \sum_{k=1}^K w_k \varrho_{\alpha_k}(Y).$ Obviously, $\varrho (Y; \varphi_d)$ is a typical example of the pessimistic risk measure in Definition 1, and it is written by
\bea
\label{eq: weighted alpha}
\varrho (Y; \varphi_d) &=& - \sum_{k=1}^K  \int_0^{\alpha_k}G(t)\frac{w_k }{\alpha_k}dt = - \int_{0}^{1} \left(\sum_{k=1}^K \frac{w_k}{\alpha_k}  I(t\leq \alpha_k)\right) G(t)dt
\eea
in terms of the $\alpha$-risk.
Because $\sum_{k=1}^K \frac{w_k}{\alpha_k}  I(t\leq \alpha_k) $
is a non-increasing function, greater weights are assigned to the lower quantiles in $\varrho(Y;\varphi_d)$.
As shown in Proposition \ref{prop: qr},  $\varrho(Y;\varphi_d)$ can be expressed in terms of the quantile risk function. 
Proposition \ref{prop: cqr} shows that the discretized version of the pessimistic risk is equal to a weighted composite quantile risk.

\begin{proposition}  \label{prop: cqr}
Assume that $Y$ is a random variable with $\E|Y| < \infty$, and $\varphi_d$ be a discrete probability measure  on $\{\alpha_k \in (0,1): \alpha_1< \cdots< \alpha_K\}$, where $\varphi_d(\{\alpha_k\}) = w_k>0$. Then  $\varrho(Y;\varphi_d) + \mu_0$ is equal to the weighted composite quantile risk of which the weights for $\alpha_k$ quantile loss are $w_k/\alpha_k$ for $k = 1, \cdots, K$.
\end{proposition}

Proposition \ref{prop: cqr} implies that the optimal portfolio minimizing the discretized pessimistic risk can be estimated by solving a composite quantile regression problem. For an arbitrary $\varphi_d$, the optimization problem minimizing  $\varrho(\bX^\top \bbeta; \varphi_d)$ with respect to $\bbeta$ is given by 
\bea
\min_{({\beta}, {\beta_0}) \in \mathbb{R}^{p}\times \mathbb{R}^K} \sum_{k=1}^K  w^\ast_k \E (\ell_{\alpha_k}(\bX^\top\bbeta - \beta_{k0})) 
\mbox{ s.t. } \hat{\bmu}^\top \bbeta = \mu_0 \mbox{ and } {\bf 1}^\top \bbeta = 1, \label{obj: weighted alpha portfolio} 
\eea
where $w^\ast_k = w_k/\alpha_k$ for $k = 1, \cdots, K$.
Conversely, when $w_k^*$s are all equal, the solution of \eqref{obj: weighted alpha portfolio} is equal to the optimal portfolio minimizing $\varrho(\cdot;\varphi_d)$ with $\alpha_k = w_k$ for all $k$. 

\section{Uniform Pessimistic Risk} \label{sec: proposal}


This section introduces the concept of Uniform Pessimistic Risk (UPR) and elaborates on its three distinct formulations: as a limit of composite quantile risk, through the lens of a proper scoring rule, and as a dual form of coherent risk. Applying a proper scoring rule facilitates the estimation of the optimal portfolio under UPR, while the dual form emphasizes the distributional robustness of the optimal portfolio. These multifaceted representations guarantee both the theoretical foundation and practical applicability of UPR, enhancing its validity as a comprehensive risk assessment tool.
\subsection{UPR as a Limit of Composite Quantile Risk}
Definition \ref{def: prm} is the general form of the pessimistic risk for a probability measure $\varphi ~\mbox{on}~ (0, 1)$. We consider the uniform distribution on $(0,1)$ as $\varphi$. Under Assumption A, UPR of a random variable $Y$ is well-defined. 

\textbf{Assumption A.} (lower tail condition) \\ 
1. There exists $r>1$ such that $\lim_{y\rightarrow -\infty} |y|^{(r+1)} f(y) = 0$, where $f$ denotes the probability density function of $Y$.  \\ 
2. The cumulative distribution function  $F$ is strictly monotonically increasing. \\ 
The assumption of lower tail condition is slightly stronger than $\E|Y|<\infty$. Therefore, the condition can be replaced by $\E|Y|^{r} <\infty$ for an arbitrary $r >1$. Further discussion will be based on the assumption of a lower tail condition of $Y$.

\begin{definition}
Let $\varphi_u$ be the uniform probability measure on $(0,1)$. UPR of $Y$ is defined by
\bean
\varrho(Y; \varphi_u) = \int^1_0 \varrho_{\alpha}(Y)d\alpha. \label{def: upr}
\eean
\end{definition}

UPR is a special case of the pessimistic risk \eqref{def: prm}, and it 
can be similarly written by an integral of the quantile function of $Y$  as \eqref{def: alpha_risk}. Proposition \ref{prop: fubini} shows that UPR is represented by a similar form as \eqref{def: alpha_risk} and its distortion function is given by $\phi(t) = - t\log t + t$  for $t\in (0,1)$. It considers the definition of $\alpha$-risk in \eqref{def: alpha_risk} in which the distortion measure is given by the uniform distribution on $(0,\alpha)$. 
\begin{proposition}\label{prop: fubini}
UPR of $Y$ is always finite. In addition, 
\bean
\varrho(Y; \varphi_u) =  - \int_0^1 G(t) d\phi(t), \mbox{ where } \phi(t) = - t\log t + t.
\eean
\end{proposition}

Assumption A is required for the finiteness of UPR. The derivative of $\phi$, $d\phi(t)=-\log t$,
\begin{wrapfigure}{r}{0.43\textwidth}
    \centering
    {\includegraphics[width=0.42\textwidth]{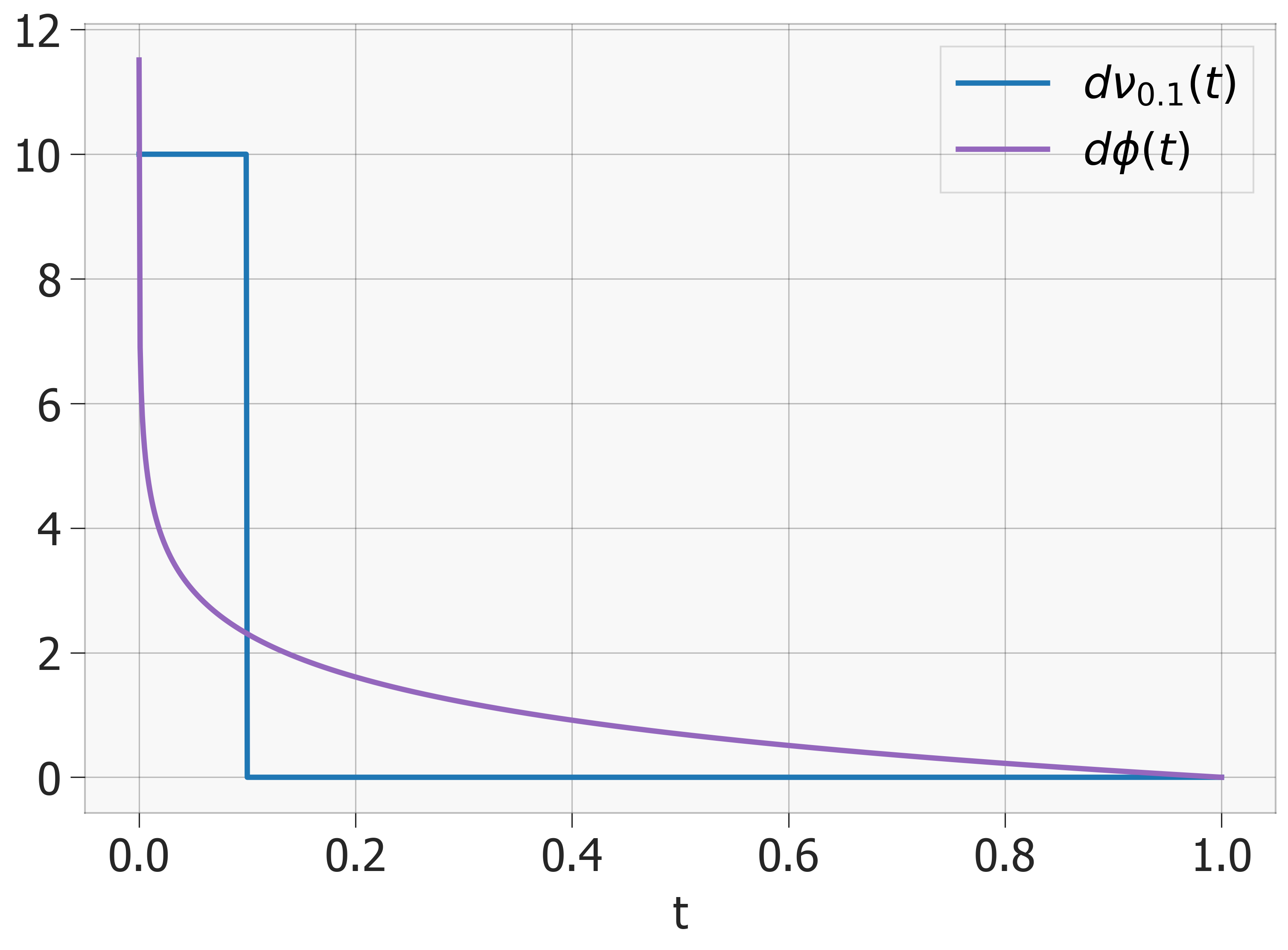}}
    \caption{Derivative functions of distortion functions. $\nu_{0.1}(t)$ and $\phi(t)$  denote the distortion functions of $\alpha$-risk at the level $10\%$ and UPR, respectively.}
    \label{fig:nu_phi}
\end{wrapfigure}
is an admissible risk spectrum \citep{acerbi2002spectral} which imposes larger weights on worse cases. 
Therefore, $\varrho(Y; \varphi_u)$ is coherent risk measure which satisfies the axioms of coherency \citep{artzner1999coherent, adam2008spectral, daouia2019extremiles} by Theorem 2.5 in \cite{acerbi2002spectral}. 

Figure \ref{fig:nu_phi} shows the shape of derivative functions of $\nu_{0.1}(t)$ and $\phi(t)$. The $\alpha$-risk at the level $10\%$ only takes the returns under 0.1-quantile into account with equal weights. However, UPR considers all returns with decaying weights.

For a continuous random variable $Y$, $\varrho_{\alpha}(Y)$ is a continuous function of $\alpha$ on $(0,1)$, which shows that UPR is the limit of the discretized pessimistic risk \eqref{eq: weighted alpha} with equal weights.

\begin{proposition} \label{approx}
Let $\varphi_{d_K}$ be a discrete uniform  probability measure  on $\{1/K, 2/K, \cdots, (K-1)/K, 1\}$.
Then, 
\bean
\lim_{K \rightarrow \infty} \varrho(Y;\varphi_{d_K}) = \varrho(Y; \varphi_u).
\eean
\end{proposition}
Through Proposition \ref{prop: cqr} and \ref{approx}, we deduce that 
the optimality induced by UPR is equivalent to the limit of composite quantile risk because 
\bea \label{scaled}
\min_{{\beta_0} \in  \mathbb{R}^K} \sum_{k=1}^K  w^\ast_k \E (\ell_{\alpha_k}(Y - \beta_{k0})) = \varrho(Y;\varphi_{d_K}) + \E(Y) \rightarrow  \varrho(Y; \varphi_u) + \E(Y),
\eea
as $K\rightarrow \infty$, where $w_k^*  = 1/k$ for $k = 1, \cdots, K$. 
\subsection{UPR as a Proper Scoring Rule}
Motivated by Proposition \ref{approx}, we can show the relationship between the quantile risk and UPR. UPR is represented as a limit of $\min_{\beta_0 \in  \mathbb{R}^K} \sum_{k=1}^K  k^{-1} \E (\ell_{\alpha_k}(Y - \beta_{k0}))$ as $K\rightarrow \infty$, where $\alpha_k = k/K$ as seen in \eqref{scaled}. 
Interestingly, UPR is closely related to the negative expected score. To elicit the relationship, we present an integral of quantile risk as a functional. Assumption B defines the domain of the functional.

\textbf{Assumption B.} (intergrable function condition) \\
Let $g$ be a continuous and non-decreasing function on $(0,1)$, and $\int_0^1 |g(\alpha)| d\alpha <\infty$. \\
Let $\mathcal{G}$ denote the set of functions that satisfy Assumption B. We define the risk functional for $g \in \mathcal{G}$ as follows:
\bea \label{eq: risk}
\mathcal{R}(g,G) &=& \int_0^1 \alpha^{-1}\E(\ell_{\alpha}(Y-g(\alpha)))d\alpha.
\eea
\begin{proposition}\label{prop: finite}
Under Assumption B, if $g$ is unbounded below, and there exists $0 < q < 1$ such that $|g(\alpha)| \asymp  \alpha^{-q}$ as $\alpha \downarrow 0$, then $0 \leq \mathcal{R}(G, G) \leq \mathcal{R}(g, G) < \infty$.
\end{proposition}
Since  $\E (\ell_{\alpha}(Y - G(\alpha))) \leq \E (\ell_{\alpha}(Y-g (\alpha)))$ for each $\alpha \in (0,1)$, $\mathcal{R}(G,G) \leq \mathcal{R}(g,G)$ for any function $g \in \mathcal{G}$. That is, we can obtain UPR of $Y$ by minimizing $\mathcal{R}(g, G)$ with respect to a function $g(\cdot)$, where $\E (Y) $ is fixed. For a $g(\cdot)$ minimizing $\mathcal{R}(g, G)$, $g(\alpha)$ is the $\alpha$-quantile of $Y$, such that the estimation of the true quantile function $G(\cdot)$ can be viewed as an extension of multiple quantiles estimation. In particular, the problem of minimizing $\mathcal{R}(g, G)$ can be addressed through the proper scoring rule. Thus, UPR of $Y$ can be computed by minimizing $\mathcal{R}(g, G)$ with respect to $g$. This methodology not only aids in quantile function estimation but also allows for an assessment of a portfolio's out-of-sample performance by comparing the function $g$ against an empirical estimate of $G$ using out-of-sample data, as detailed by \citep{emmer2015best}. Visual examples of out-of-sample performance evaluation are available in the Appendix.  

We adopt the method from \cite{gasthaus2019probabilistic} which parameterizes the function $g \in \mathcal{G}$ using a linear isotonic regression spline, as follows: 
\bea 
g_{\theta}(\alpha) = \gamma + \sum_{m=0}^M b_m(\alpha - d_m)_+ \label{eq: lirs}
\mbox{ s.t. } \sum_{m=0}^{j} b_{m} > 0, ~ j = 0, \dots, M,
\eea
where $\theta = (\gamma, {\bf b})$, $\gamma \in \mathbb{R}$, ${\bf b} = (b_0, \dots, b_M)\in \mathbb{R}^{M+1}$, ${\bf d} = (d_0, \dots, d_M)\in [0, 1]^{M+1}$, $0 = d_0 < \cdots  < d_{M} = 1$,  and $(z)_+ = \max(0, z)$. The constraints of $b_m$ guarantee the monotonically increasing property of $g_\theta$ on $(0, 1)$. In our paper, the knot points are fixed for the convexity of the objective function \eqref{eq: risk}. \\ 
Unfortunately, when $g$ is bounded below, then $\mathcal{R}(g,G)$ is infinite. 
Proposition \ref{prop: infty} shows that $\mathcal{R}(g, G)$ is infinite for $g$, which is bounded below when $G$ is unbounded.
\begin{proposition} \label{prop: infty}
Suppose that $G$ is an unbounded quantile function of a continuous distribution. If $g$ is bounded below, then $\mathcal{R}(g, G) = \infty.$
\end{proposition}
To estimate the true quantile function $G$, the empirical risk for $\mathcal{R}(g, G)$ can be used as an objective function. In the presence of the unboundedness of $\mathcal{R}(g, G)$, the approximation of the empirical risk cannot be theoretically guaranteed, and the instability of computation increases in the middle of estimating the quantile function. 
To resolve this problem, we define  $\mathcal{R}_\eta (g, G)$ for $\eta>0$ as follows:
\bea \label{eq: eta_risk}
\mathcal{R}_\eta (g, G) = \int_\eta^1 \alpha^{-1} \E (\ell_\alpha (Y - g(\alpha)))d\alpha.
\eea
Proposition \ref{prop: finite2} shows that $\mathcal{R}_\eta (g, G)$ always has a finite value under Assumption B. 
\begin{proposition} \label{prop: finite2} Under Assumption B, let $G$ be a quantile function of $Y$ and  $g \in \mathcal{G}$ be bounded below function. For $\eta \in (0,1)$, \\ 
$$0 \leq \mathcal{R}_\eta (G, G)\leq \mathcal{R}_\eta (g, G) < \infty
\mbox{ and } \lim_{\eta \downarrow 0} R_{\eta}(G, G) = R(G,G).$$
\end{proposition} 
For observations $y_1,\cdots, y_n$ from $F$, a natural approximation of $\mathcal{R}_\eta(g_{\theta}, G)$ is 
$$\mathcal{R}_\eta^n(g_{\theta}) = \frac{1}{n}\sum_{i=1}^n \int_\eta^1 \alpha^{-1} \ell_{\alpha}(y_i -g_{\theta}(\alpha))d\alpha.$$ 
Theorem \ref{thm: lln} shows that $\mathcal{R}_\eta^n(g_{\theta})$ is an empirical version of $\mathcal{R}_\eta(g_{\theta},G)$.
\begin{theorem} \label{thm: lln}
Suppose that $y_i$ for $i = 1, \cdots, n$ are random samples of $Y$.  
Let $\mathcal{R}_\eta^n(g_{\theta}) = \frac{1}{n}\sum_{i=1}^n \int_\eta^1 \alpha^{-1} \ell_{\alpha}(y_i -g_{\theta}(\alpha))d\alpha$, then 
$\mathcal{R}_\eta^n(g_{\theta}) \rightarrow \mathcal{R}_\eta(g_{\theta},G)$ with probability 1 as $n\rightarrow \infty$ for each $\theta$.
\end{theorem}

In addition, the M-estimator minimizing $\mathcal{R}_\eta^n(g_{\theta})$ converges to a minimizer of $\mathcal{R}_\eta(g_{\theta},G)$. 
If the parameter space of $\theta$ is restricted to a compact space on which $\mathcal{R}_\eta^n(g_{\theta}, G)$ is continuous at $\theta$ for almost surely, then the result of Wald's consistency implies the convergence of the M-estimator to $\theta_0$ minimizing $\mathcal{R}_\eta^n(g_{\theta}, G)$. 

\begin{theorem} \label{thm: consistency}
 Let $\Theta$ be a compact subset of $\{(\gamma, {\bf b})\in \mathbb{R} \times \mathbb{R}^{M+1}:
\sum_{m=0}^{j} b_{m} \geq 0, ~ j = 0, \dots, M\}$ satisfying the following conditions:\\ 
1. For each $\theta\in \Theta$ there exist $\omega:[\eta,1) \rightarrow \mathbb{R}$ and a small neighborhood $U$ of $\theta$, such that $\sup_{\theta \in U}\frac{1}{\alpha}|g_{\theta}(\alpha)| \leq  \omega(\alpha)$ and $\int_\eta^1 |\omega(\alpha)| d\alpha <\infty$.\\ 
2. For a sequence $\theta_n\in \Theta$ goes to $\theta$, a pointwise limit $ g_{\theta}(\alpha) = \lim_{\theta_n \rightarrow \theta} g_{\theta_n}(\alpha)$ exists.\\ 
Let $\Theta_0 = \{\theta_0 \in \Theta: \mathcal{R}_{\eta}(g_{\theta_0},G) = \inf_{\theta\in \Theta}\mathcal{R}_{\eta}(g_{\theta}, G)\}$ and suppose that we have $\hat \theta_n$ satisfying
$\mathcal{R}_\eta^n(g_{\hat{\theta}_n}) \leq \mathcal{R}_\eta^n(g_{\theta_0}) - o_p(1)$ for $\theta_0 \in \Theta_0$. Then, 
\bean
\Pr\left(d(\hat \theta_n, \Theta_0) >\epsilon,  \hat \theta_n \in \Theta \right) \rightarrow 0 \mbox{ as } n\rightarrow \infty.
\eean

\end{theorem}

\subsection{UPR as a Law Invariant Coherent Risk Measure} \label{sec: robust}
\cite{kusuoka2001law} showed that for a bounded random variable $Y$, the law invariant coherent risk
measure $\varrho(Y)$ is represented by a term of $\alpha$-risk as follows:
\bea \label{eq: law inv}
\varrho(Y)= \sup_{\varphi \in \mathcal{P}} \int_0^1 \varrho_{\alpha}(Y) d\varphi(\alpha),
\eea
where $\mathcal{P}$, referred to as risk envelope, denotes a set of probability measures on the interval $(0, 1)$.
Intuitively, the right term of \eqref{eq: law inv} can be thought of as the worst-case pessimistic risk under $\varphi \in 
\mathcal{P}$. \cite{shapiro2012minimax} and \cite{ruszczynski2006optimization} generalize \eqref{eq: law inv} for a random variable $Y$ in $L_p$ space with $p \in [1,\infty]$.  
Because the $\alpha$-risk is coherent and UPR is a nonnegative weighted sum, UPR is also coherent. Thus, UPR can be written by a worst-case pessimistic risk, as in Proposition \ref{prop: dro}.

\begin{proposition} \label{prop: dro}
Let $Y \in L_p$ for $p \geq 1$  and $Z \in L_q$ with $1/p + 1/q = 1$. Denote the cumulative distribution function of $Z$ by $Q$ and the class of all cumulative distribution functions associated with $Z\in L_q$ by  $\mathfrak{M}_q$. Then, there exists a convex subset $\mathcal{M}_{u} \subset \mathfrak{M}_q$ such that 
\[
\varrho(Y; \varphi_u) = \sup_{Q \in \mathcal{M}_{u}} -\E_{Q}(Z),
\mbox{ where }
\mathcal{M}_{u} = \{ Q \in \mathfrak{M}_q ~ | ~ -\int_0^1 Q^{-1}(t) dt \leq \varrho(Y;\varphi_u)  \}.
\]
\end{proposition}

According to Proposition \ref{prop: dro}, UPR is equivalent to worst-case expectation under all probability measures in $\mathcal{M}_u$. When the worst-case expectation is applied to define a risk function, the problem of minimizing the risk is known as DRO \citep{rahimian2019distributionally}. If $Z \sim Q$ denotes an asset's distributionally shifted return, $-\E_Q(Z)$ represents the expected loss under $Q$, then Proposition \ref{prop: dro} implies that $\varrho(Y; \varphi)$ is the maximum expected loss among a distribution class $\mathcal{M}_u$. Thus, the optimality based on UPR implies minimizing the worst-case expected loss under a suitable distribution class. In the next section, we define the distribution class in which UPR is the worst-case expected loss employing the beta distribution.

\begin{remark}
The coherent risk presented in \cite{ruszczynski2006optimization} measures the risk associated with the loss of an asset. If $Y$ is the negative loss, then Proposition \ref{prop: dro} is consistent with the result of Corollary 1 \citep{ruszczynski2006optimization}.
\end{remark}

\subsection{Extension of UPR} 

UPR is naturally generalized by employing a different distortion measure instead of a uniform distribution. Here, we consider a class of beta distributions that includes the uniform distribution. 
Denote the density function of the beta distribution with parameters $(s,h)$ by $b(\cdot; s,h)$, and define the family of the pessimistic risk by 
\bea \label{eq: beta}
\varrho_{(s,h)}(Y) = \int_0^1  \varrho_{\alpha}(Y) b(\alpha;s,h)d\alpha,
\eea
for $s > 0$ and $h >0$. 
Further, \eqref{eq: beta} can be rewritten in terms of the quantile function of $Y$ and the associated distortion function as $\varrho_{(s,h)}(Y) = - \int_0^1 G(t) \phi(t;s,h) dt$, where $\phi(t;s,h) = \frac{B(1, s-1, h) - B(t, s-1, h)}{B(1,s,h)}$ and 
$B(x, s, h) = \int_0^x \alpha^{s-1}(1-\alpha)^{h-1} d\alpha$.

For a fixed $Y$ with $Y\in L_p$, the value of $\varrho_{(s,h)}(Y)$ determines the class of shifted distributions presented in Proposition \ref{prop: dro}, $\mathcal{M}_{(s,h)} = \{Q \in \mathfrak{M}_q| -\int_0^1 Q^{-1}(t) dt \leq \varrho_{(s,h)}(Y)\},$ where $1/p + 1/q = 1$. Because  $\mathcal{M}_{(s,h)}$ depends on  $\varrho_{(s,h)}(Y)$, the distributional robustness can be modulated by choice of $s$ and $h$.
Especially, when $Y$ follows the generalized extreme value (GEV) distribution \citep{coles2001introduction},  we can show that $\varrho_{(s,1)}(Y)$ is a decreasing function  of $s\geq 1$. That is, for the special case of $Y$, we verify that $\varrho(Y;\varphi_u) = \sup_{s\geq 1}\varrho_{(s,1)}(Y)$, which implies that UPR is the most conservative risk against distribution shift. Theorem \ref{thm: gev} introduces the example that the generalized risk $\varrho_{(s,1)}(Y)$ are nested with respect to their associated shifted distributions. 
\begin{theorem} \label{thm: gev}
Let $Y$ be the random variable following the generalized extreme value distribution. If $\E|Y|<\infty$, then 
$\varrho_{(s,1)}(Y)$ is non-decreasing function of $s\geq 1$, i.e., $\mathcal{M}_{(s,1)} \subset \mathcal{M}_{(1,1)}$ for all $s\geq 1$. 
\end{theorem}
Another advantage of employing the beta distribution is the convergence of the risk functional \eqref{eq: risk}, even when the function $g$ is bounded below. For convenience and alignment with a pessimistic perspective, we focus solely on the beta distribution where $h \geq 1$. Proposition \ref{prop: s-cond} demonstrates the conditions under which the beta distribution ensures a finite risk functional, regardless of whether $g$ is bounded below.
\begin{proposition} \label{prop: s-cond}
Let $b(\alpha; s, h)$ be a density function of the beta distribution with parameters $s, h \geq 1$. Assume that a continuous and non-decreasing function $g$ satisfies Assumption B. 
If $g$ is not bounded below, suppose that there exists $0 < q < 1$ such that $|g(\alpha)| \asymp  \alpha^{-q}$ as $\alpha \downarrow 0$, then  
\bean
\int_0^1 \frac{1}{\alpha} \E(\ell_\alpha (Y - g(\alpha))) b(\alpha; s, h) d\alpha < \infty.
\eean
However, if $g$ is bounded below, the above condition for finiteness holds when $s>1$.
\end{proposition}
\begin{remark}
If the pessimistic risk is defined by $b(\alpha; 2, 1)$, i.e. $\varrho_{(2,1)}(Y)$, its risk functional of $g$ is formed by the continuous ranked probability score (CPRS) \citep{gneiting2007strictly, gneiting2011comparing}, which is the most well-known example of the proper scoring rule as follows:
\bean
\int_0^1 2 \E(\ell_{\alpha}(Y-g(\alpha))) d\alpha = \E\left(\int_0^1 2\ell_{\alpha} (Y- g(\alpha))d\alpha \right) 
= \E \left[CRPS(Y, g(\alpha))\right].
\eean
For $Y$ with $\E|Y|<\infty$, $\E[CRPS(Y, g)]$ is finite. By Theorem \ref{thm: gev}, UPR is more conservative than CRPS when the underlying distribution of $Y$ is the GEV distribution. 
\end{remark} 

\section{Optimal portfolio with UPR} \label{sec: portfolio}
Section \ref{sec: portfolio} explains the minimization problem to obtain the optimal portfolio based on UPR. We consider $p$ as the number of underlying assets and denote the vector of their log returns by $\bX \in \mathbb{R}^p$. Then, the return of a portfolio with a weight $\bbeta \in \mathbb{R}^p$ is given by $Y = \bX^\top \bbeta$ with  ${\bf 1}^\top \bbeta = 1$.

Let the expected return on $p$ assets be $\E(\bX) = \bmu$ and denote UPR of the portfolio $\bX^\top \bbeta$ by  $\varrho(X^\top \bbeta; \varphi_u)$. 
$G_{\beta}$ denotes the quantile function of $\bX^\top \bbeta$. UPR of the portfolio $\bX^\top \bbeta$ is written by the infimum of the risk functional, $\inf_{g \in \mathcal {G}}  \mathcal{R}_\eta(g, G_{\beta})$ defined in \eqref{eq: eta_risk}. Abusing notation, denote $\mathcal{R}_\eta(g, G_{\beta})$ by $\mathcal{R}_\eta(g, \bbeta)$ and let $\mathcal{R}_\eta^n(g, \bbeta) =  \frac{1}{n} \sum_{i=1}^n \int_\eta^1 \alpha^{-1} \ell_{\alpha}( \bx_i^\top \bbeta - g(\alpha))d\alpha$, where $\bx_i$s are random samples of $\bX$.

For a fixed target return $\E(\bX^\top \bbeta) = \mu_0$, 
we can estimate an optimal portfolio weight by  minimizing $\mathcal{R}_\eta^n(g, \bbeta)$, an empirical analog of $\mathcal{R}_\eta(g,\bbeta)$.  
\bean
\min_{(\beta,g) \in \mathbb{R}^p \times \mathcal{G}} \mathcal{R}_\eta^n(g, \bbeta) \mbox{ s.t. }  \hat{\bmu}^\top \bbeta = \mu_0 \mbox{ and } {\bf 1}^\top \bbeta = 1, \nonumber 
\eean
where $\hat{\bmu}$ denotes the sample mean vector of $\bx_i$s. \\ 
Finally, an optimal portfolio with UPR can be obtained by solving the following problem:
\bea
\min_{(\beta, \theta) \in \mathbb{R}^{p+M+2}}  \mathcal{R}_\eta^n(g_\theta, \bbeta) \label{obj: sample upr with lirs} 
\mbox{ s.t. }  \hat{\bmu}^\top \bbeta = \mu_0 \mbox{ and } {\bf 1}^\top \bbeta = 1, \sum_{m=0}^{j} b_{m} > 0, ~ j = 0, \dots, M.
\eea
The optimal portfolio minimizing UPR is the solution to the convex problem \eqref{obj: sample upr with lirs}. Let $(\hat{\bbeta}, \hat{\theta})$ be the solution of the problem \eqref{obj: sample upr with lirs}, then $\hat{\bbeta}$ is the optimal portfolio weight and $g_{\hat{\theta}}(\cdot)$ is the estimated quantile function of the portfolio return. Theorem \ref{thm: beta_consistency}, the extended version of Theorem \ref{thm: consistency}, shows the optimal parameters of \eqref{obj: sample upr with lirs} can be consistently estimated. 
\begin{theorem} \label{thm: beta_consistency}
Let $\psi =(\bbeta, \theta)$, and assume that the following conditions hold: \\
1. Let a parameter space $\Psi$ be a compact subset of $\{\psi = (\bbeta, \gamma, {\mathbf b}) \in \mathbb{R}^{p+M+2}: \sum_{m=0}^j b_m \geq 0, j = 1, \dots, M \}$. \\
2. $\bX$ is $p$-dimensional vector of $\sigma^2$-sub Gaussian random variables, and $\bx$ is the realization of $\bX$.\\
3. Let $\psi_0 = (\bbeta_0, \gamma_0, \mathbf{b}_0) \in \Psi$ such that $ \mathcal{R}_{\eta}(g_{\theta_0},\bbeta_0) = \underset{\psi \in \Psi} {\inf}\mathcal{R}_{\eta}(g_{\theta}, \bbeta)$, where $\theta_0 = (\gamma_0, \mathbf{b}_0)$, and we have $\hat \psi_n = (\hat{\bbeta}_n, \hat{\theta}_n)$ satisfying $\mathcal{R}_\eta^n(g_{\hat{\theta}_n}, \hat{\bbeta}_n) \leq \mathcal{R}_\eta^n(g_{\theta_0}, \bbeta_0) + o_p(1)$. Then, $\hat \psi_n \overset{p}{\rightarrow} \psi_0.$
\end{theorem}
Through Proposition \ref{prop: finite2}, the parametrization of $g$ with $\theta$ in \eqref{eq: lirs} yields a closed form for $\mathcal{R}_\eta^n(g_\theta,\bbeta)$ as follows:
\bean
\mathcal{R}_\eta^n(g_\theta,\bbeta) &=& \frac{1}{n}\sum_{i=1}^n \bigg(\left(1-\eta+\log\tilde{\alpha}_i\right)\left(\bx_i^{\top}\bbeta-\gamma\right) \\ 
&+& \sum_{m=0}^M b_m\bigg(1-\frac{(1-d_m)^2}{2} - \max(\tilde{\alpha}_i, d_m) + \max(\log \tilde{\alpha}_i, \log d_m)d_m\bigg)\bigg),
\eean
where $\tilde{\alpha}_i =  g^{-1}_\theta(\bx_i^{\top}\bbeta) = \frac{\bx_i^{\top}\bbeta - \gamma + \sum_{m=0}^{m_i} b_m d_m}{\sum_{m=0}^{m_i} b_m}$ with $d_{m_i} \leq \tilde{\alpha}_i < d_{m_{i+1}}$, and $m_{i}$ is the largest knot position index such that $g_\theta(d_{m_i}) \leq g_\theta(\tilde{\alpha}_i)$.

For ease of implementation, we reparametrize $\bf b$ as described in \cite{gasthaus2019probabilistic} and adopt the two-step alternating updating rules of $\bbeta$ and $\theta = (\gamma, \mbox{\bf b})$. Here, we reparameterize $ \mbox{\bf b} $ as $\bdelta$, where 
$b_0 = \delta_0 > 0$ and $b_m = \delta_m - \delta_{m-1}$ for $m = 1, \cdots, M$. The inequality constraints in \eqref{obj: sample upr with lirs} are rewritten as $\delta_{m} \geq 0, ~ m = 1, \dots, M$. First, $\bbeta$ and $(\gamma,\bdelta)$ are updated by the gradient descent methods, and $\bdelta$ is projected onto feasible set, $\{\delta_m; \delta_m > 0, ~ m=0, \dots, M\}$. Next, the projected gradient method is applied to update $\bbeta$ with the constraints of $\hat{\bmu}^\top \bbeta = \mu_0$ and ${\bf 1}^\top \bbeta = 1$. 
Our algorithm utilizes a gradient descent approach. Additionally, the closed-form expression of the UPR requires an ordering process for the knot index where each sample resides. This ordering is proportional to the sample size and the number of knot points, reflecting the complexity of the spline models. Therefore, the computational complexity of our algorithm can be described as $O(p+n(M+1))$. All the steps for our optimal UPR portfolio are described in the Appendix.

\section{Numerical Studies} \label{sec: num}
In this section, we present numerical results using three real stock datasets: S\&P500, CSI500, and KOSPI200. 
In the real data analysis, the out-of-sample performances are evaluated with the portfolio returns of the stock datasets. The experiments were conducted using \textsf{Tensorflow} with \textsf{Python}, and the source code is accessible at XXX. We set $\eta=10^{-5}$ in our implementation. In the Appendix, we investigate the properties of UPR, particularly in scenarios prone to extreme risks with the synthetic data, by modeling tail dependencies with the Clayton copulas.  

\subsection{Experimental Setup}
We utilize the three stock datasets, consisting of daily log return rates spanning from 2013 to 2021. These log return rates are calculated using the adjusted closing stock prices. To assess the out-of-sample performance, we adopt a rolling sample estimation strategy \citep{demiguel2009optimal}. Each optimal portfolio is determined based on the daily log return rates from the preceding 240 days, and the daily returns of this optimal portfolio are then computed over the subsequent 60 days. Following this, the estimation and evaluation time windows are advanced by another 60 days, and the estimation and evaluation process is iteratively repeated.

\subsection{Benchmark Models and Metrics} 
We evaluate our proposed portfolio model, UPR, against seven benchmark portfolios: an equally-weighted portfolio (EW) that invests in equal proportions in all assets, the quantile regression model (QR) with $\alpha=0.1$ in \eqref{eq: qr_model} (also called mean-CVaR portfolio \cite{rockafellar2000optimization}),
two composite quantile regression models (CQR1 and CQR2) considering three quantile levels $\{0.1, 0.5, 0.9\}$ and $\{0.01, 0.1, 0.5, 0.9\}$ with weight vector $(1/3, 1/3, 1/3)$ and $(0.4, 0.3, 0.2, 0.1)$ in \eqref{eq: weighted alpha}, a linear-programing-based CVaR portfolio models (EO) and its DRO version (RO) \cite{chen2022robust}, and the mean-variance portfolio (MV). For details of benchmark models, please refer to the Appendix. 

We use five performance metrics to evaluate benchmark models: cumulative wealth (CW), maximum drawdown (MDD), maximum loss (MaxLoss),  CVaR at level $0.1$ (CVaR$_{0.1}$), and Sharpe ratio (SR). For details of evaluation metrics, please refer to the Appendix.     

\subsection{Performance Results} 
\begin{figure}[t]
    \centering
    \subfigure[S\&P500]{
    \includegraphics[width=0.32\textwidth]{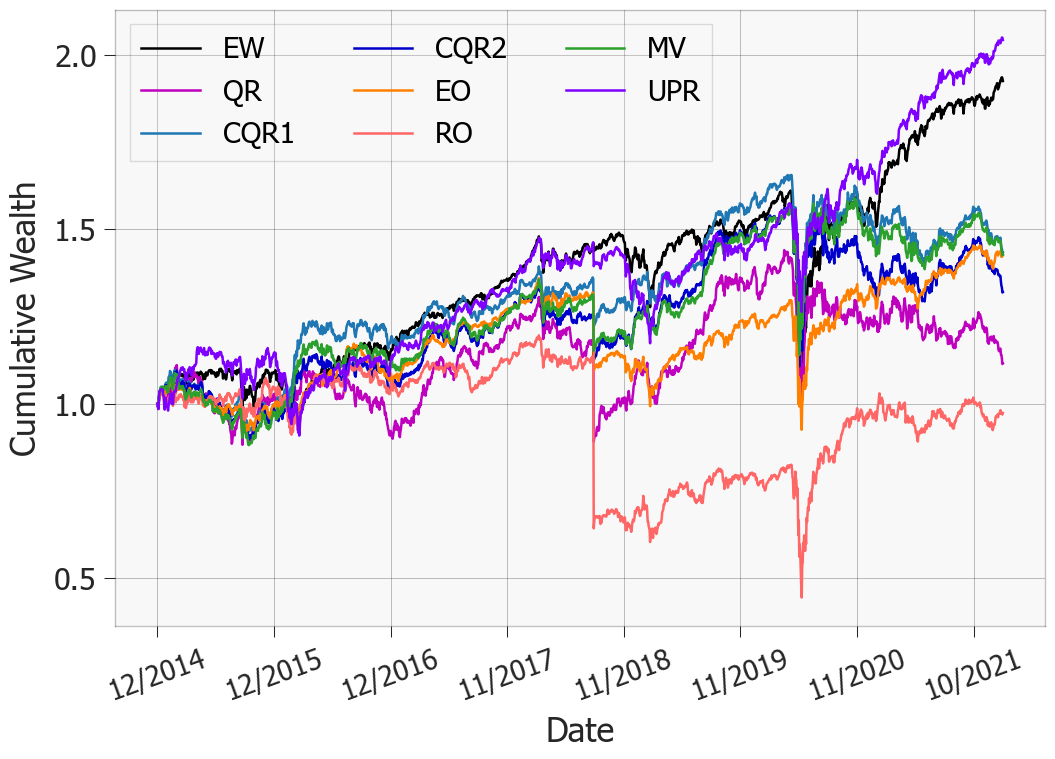}}
    \subfigure[CSI500]{
    \includegraphics[width=0.32\textwidth]{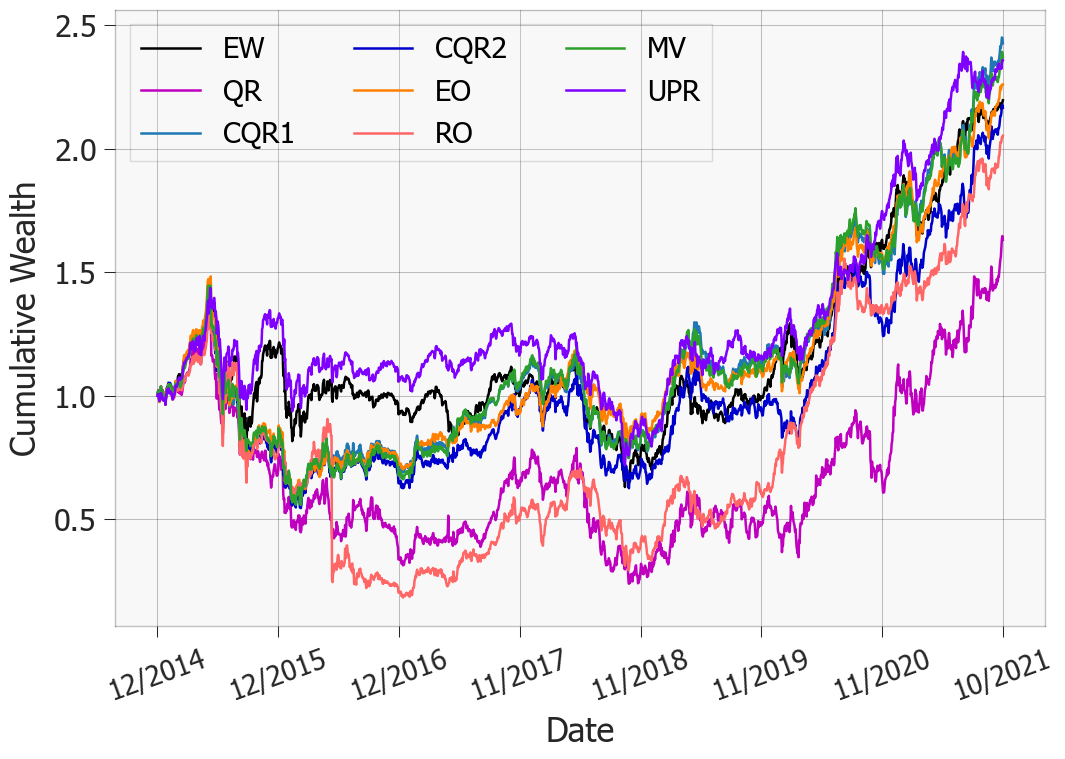}}
    \subfigure[KOSPI200]{
    \includegraphics[width=0.32\textwidth]{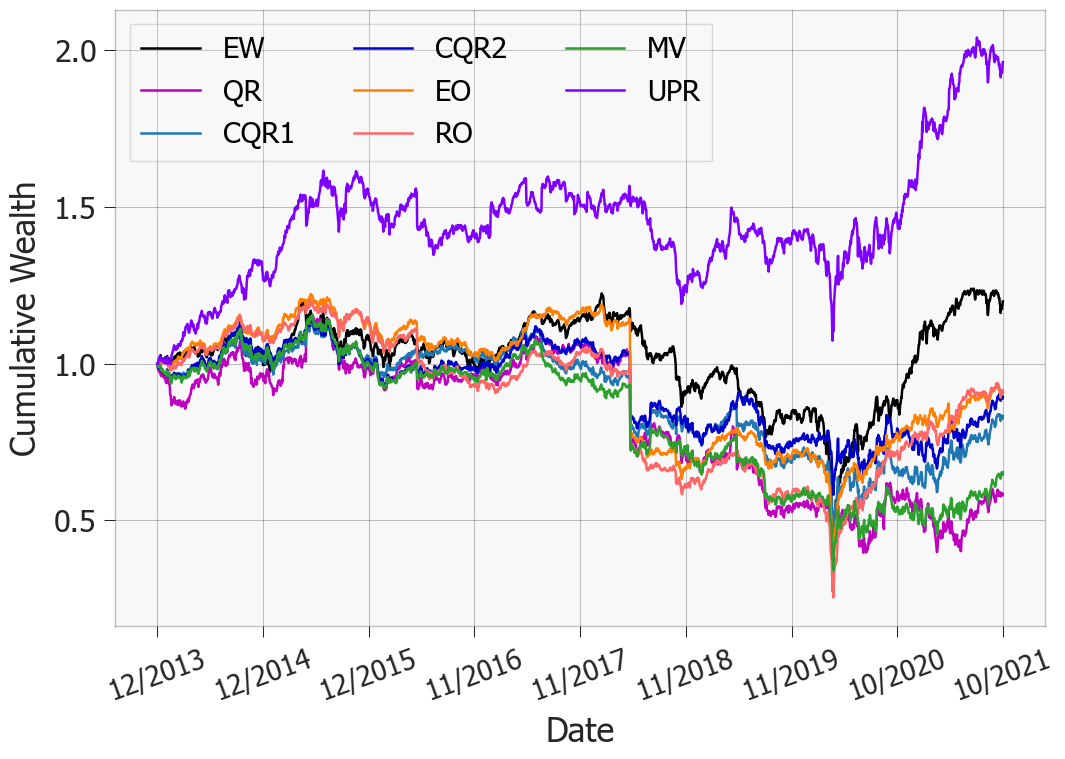}}
    \caption{Cumulative wealth of the portfolio models.}
    \label{fig:cw}
\end{figure}

Table \ref{tab:result_rda} showcases the performance metrics over the total evaluation periods, with the out-of-sample results distinctly underscoring the advantages of our proposed risk and portfolio model. Particularly in terms of the rate of return, evidenced by CW and SR metrics, UPR uniquely outperforms the EW benchmark, indicative of market return excluding the CSI500 dataset. We also conduct statistical tests of SR \cite{memmel2003performance, demiguel2009optimal} between the UPR portfolio models and the others. Figure \ref{fig:cw} illustrates the CW of all portfolio models during the total evaluation periods. MaxLoss and MDD, which represent the extreme losses of concern to investors, exhibit the most favorable values in UPR. We argue that the superiority and robustness of UPR over other pessimistic risks pertain to the distortion function $\phi$. 
Considering entire returns with decaying weights gives rise to robust out-of-sample performance. 
Furthermore, it is noteworthy that both UPR and CQRs consistently outperform the DRO version of CVaR models across all datasets. This suggests that incorporating information from multiple quantiles, rather than relying solely on a single quantile, enhances portfolio robustness even in the absence of DRO techniques. In the Appendix, providing the quantile discrepancy between in-sample and out-of-sample of UPR and CQRs, the entire quantile estimation which reflects the most conservative return distribution and our DRO property are more adequate for robust portfolio optimization.

\begin{table}[t]
    \centering
    \caption{Out-of-performance measures of the portfolio models. The most favorable value is bolded. $^*$ means that there is a significant difference between UPR and the comparison model in SR. }
    \begin{adjustbox}{width=\textwidth}
    \begin{tabular}{l c c c c c c c c c c c c c c c} 
    \Xhline{1.1pt}  
         & \multicolumn{5}{c}{S\&P500} & \multicolumn{5}{c}{CSI500} & \multicolumn{5}{c}{KOSPI200} \\ 
        \cmidrule(lr){2-6} \cmidrule(lr){7-11} \cmidrule(lr){12-16}
        Model & CW$\uparrow$ & MaxLoss$\downarrow$ & MDD$\uparrow$ & CVaR$_{0.1} \downarrow$ & SR$\uparrow$ & CW$\uparrow$ & MaxLoss$\downarrow$ & MDD$\uparrow$ & CVaR$_{0.1} \downarrow$ & SR$\uparrow$ & CW$\uparrow$ & MaxLoss$\downarrow$ & MDD$\uparrow$ & CVaR$_{0.1} \downarrow$ & SR$\uparrow$ \\ \hline 
      EW        & 1.925  & 0.134 & -0.319 & 0.0218 & 0.044 & 2.196  & 0.091 & -0.552 & 0.0369 & 0.038 & 1.199  & 0.121 & -0.722 & 0.0196 & 0.010$^*$ \\
      QR        & 1.166  & 0.274 & -0.317 & 0.0248 & 0.005$^*$ & 1.631  & 0.224 & -0.827 & 0.0418 & 0.017 & 0.583 & 0.274 & -0.658 & 0.0208 & -0.017$^*$ \\
      CQR1      & 1.434  & 0.135 & -0.267 & 0.0203 & 0.021$^*$ & \textbf{2.426}  & 0.180 & -0.617 & 0.0343 & \textbf{0.045} & 0.833 & 0.162 & -0.613 & 0.0172 &  -0.009$^*$  \\
      CQR2      & 1.320  & 0.133 & -0.292 & 0.0215 & 0.014$^*$ & 2.165  & 0.151 & -0.612 & 0.0352 & 0.036 & 0.894 & 0.209 & -0.459 & 0.0169 &  -0.005$^*$  \\
      EO      & 1.427  & 0.223 & -0.320 & \textbf{0.0183} & 0.022$^*$ & 2.260  & 0.155 & -0.608 & \textbf{0.0341} & 0.042 & 0.820 & 0.255 & -0.817 & 0.0170 &  -0.009$^*$  \\
      RO      & 0.974  & 0.480 & -0.628 & 0.0212 & -0.001$^*$ & 2.052  & 0.582 & -0.866 & 0.0385 & 0.026 & 0.747 & 0.243 & -0.882 & \textbf{0.0158} &  -0.014$^*$  \\
      MV        & 1.427  & 0.156 & -0.271 & 0.0203 & 0.020$^*$ & 2.371  & 0.164 & -0.614 & 0.0342 & 0.043 & 0.653 & 0.196 & -0.706 & 0.0171 &  -0.018$^*$ \\
      UPR       & \textbf{2.043}  & \textbf{0.115} & \textbf{-0.218} & 0.0209 & \textbf{0.051} & 2.358  & \textbf{0.089} & \textbf{-0.499} & 0.0372 & 0.042 & \textbf{1.962}  & \textbf{0.108} & \textbf{-0.336} & 0.0227 & \textbf{0.038} \\
    \Xhline{1.1pt}   
    \end{tabular}
    \end{adjustbox}
    \label{tab:result_rda}
\end{table}

\section{Conclusion and limitations} \label{sec: conc}
This study extensively analyzed various representations of UPR for theoretical foundations. Moreover, we introduced a computational algorithm designed to evaluate the proposed risk for its various applications. To the best of our knowledge, we first propose the optimization algorithm with the integral of $\alpha$-risk with the gradient descent method.
Additionally, the numerical studies revealed that UPR outperforms other methods. These findings suggest that UPR could be an effective alternative in domains like risk-averse reinforcement learning \cite{singh2020improving} and supervised learning \cite{leqi2022supervised}, potentially replacing $\alpha$-risk or CVaR.

Our work leaves some open questions for future research. 
The optimization of UPR depends on the linear spline model $g_\theta$. The extension of $g_\theta$ through non-parametric or deep learning approaches \cite{liu2022kernel, narayan2024expected} would be of interest. In addition, we extend the probability measure $\varphi$ in \eqref{def: prm} to other distributions for optimal selection of the distortion function. One possible solution is the extension of the distortion measures with the beta mixture distribution \cite{rousseau2010rates}. We leave them as our future works.

\bibliography{upr}

\appendix

\section{Proofs}
\subsection{Proof of Proposition 1}
\begin{proof}
 Let $\ell_{\alpha}(z) = (\alpha - I(z<0))z$, then
\bea \label{eq:eqrisk1} 
\E(\ell_{\alpha} (Y - \beta_0))  &=&  \E( (\alpha - I(Y - \beta_0<0))(Y - \beta_0)) \nonumber
\\
&=& \alpha(\E(Y)-\beta_0) - \int^{\beta_0}_{-\infty}(y-\beta_0)dF(y) \nonumber\\
&=& \alpha \E(Y) - \alpha \beta_0 - \int_{-\infty}^{\beta_0} ydF(y) + \beta_0 F(\beta_0) \nonumber \\
&=& \alpha \E(Y) - \int_0^{F(\beta_0)} G(t)dt  + \beta_0( F(\beta_0)- \alpha).
\eea
Because $\underset{\beta_0}{\mbox{min}} ~ \E(\ell_\alpha (Y -\beta_0)) = \E(\ell_\alpha (Y -G(\alpha))$, plugging $G(\alpha)$ into $\beta_0$ in \eqref{eq:eqrisk1} leads to 
\bean 
\underset{\beta_0}{\min} ~ \E(\ell_\alpha (Y - \beta_0)) 
&=& \alpha \E(Y) - \int_0^\alpha G(t)dt \nonumber \\ 
&=& \alpha (\mu_0 + \varrho_{\alpha} (Y)).
\eean
\end{proof}

\subsection{Proof of Proposition 2} 
\begin{proof}
For given $\{\alpha_k\}_{k=1,\dots,K}$, let  $ \beta_{k0}^*$ be the minimizer of $\E (\ell_{\alpha_k}(Y-\beta_{k0}))$ with respect to $\beta_{k0}$ for $k = 1, \cdots, K.$ Then, 
\bean
\varrho(Y;\varphi_d) +\E(Y) &=& \sum_{k=1}^K w_k\left(\varrho_{\nu_{\alpha_k}}(Y) + \E(Y)\right)\\ &=&\sum_{k=1}^K \frac{w_k}{\alpha_k}  \E(\ell_{\alpha_k}(Y - {\beta}_{k0}^*))  \\ 
&=& \sum_{k=1}^K w^\ast_k \E(\ell_{\alpha_k}(Y - {\beta}_{k0}^*))
\eean 
where $w^\ast_k = w_k/\alpha_k$. The second equality is held by Proposition \ref{prop: qr}.
\end{proof}

\subsection{Proof of Proposition 3} 
\begin{proof}
Denote the probability density function of $Y$ by $f(\cdot)$.
By change of variables with $G(\alpha)=t$, we can rewrite $\varrho(Y; \varphi_u)$ as follows:
\bean 
\varrho(Y; \varphi_u) &=& - \int_0^1 \frac{1}{\alpha}\int_0^\alpha G(t) dt d\alpha\\
&=& - \int_0^1 \int_{-\infty}^{G(\alpha)} \frac{1}{\alpha} y f(y) dy d\alpha \\
&=& - \int_{-\infty}^{\infty} \int_{-\infty}^t \frac{1}{F(t)} yf(y)f(t) dy dt \\ 
&=& - \int_{-\infty}^\infty y f(y) \left[\log F(t)\right]_{t=y}^{\infty} dy = \int_{-\infty}^\infty yf(y)\log F(y) dy.
\eean
By Fubini's theorem, the third equality above holds whenever $\int_{-\infty}^\infty | yf(y)\log F(y) |dy < \infty$, which is guaranteed by  Assumption A. Thus,  the exchange of double integrals in $\varrho(Y; \varphi_u)$  leads to 
\bean
\varrho(Y; \varphi_u) &=& -\int^1_0 \frac{1}{\alpha}\int_0^\alpha G(t) dt d\alpha \nonumber \\
&=& - \int_0^1 G(t) \int_t^1 \frac{1}{\alpha} d\alpha dt \\
&=& - \int_0^1 G(t) (-\log t) dt \label{eq:ars form}.
\eean
\end{proof}

\subsection{Proof of Proposition 5}
\begin{proof}
$\mathcal{R}(G, G)\geq 0$ is trivial because $\ell_{\alpha}(Y-G(\alpha)) \geq 0$ almost surely. Proposition 1 implies that $\int_0^1 \left(\varrho_{\alpha}(Y) + \E(Y) \right) d\alpha =  \int_0^1 \alpha^{-1} \E  ( \ell_{\alpha}(Y- G(\alpha)))  d\alpha = \mathcal{R}(G, G)$, which concludes $\mathcal{R}(G, G)<\infty$ by Proposition 3. 

From Proposition 1, we have $\E[\ell_\alpha (Y - G(\alpha))] \leq \E[\ell_\alpha (Y - g(\alpha))]$ for any $g$. Therefore,
\bean
&& \mathcal{R}(G,G) - \mathcal{R}(g, G) \nonumber \\
&=&\int_0^1 \left( \frac{1}{\alpha} \E\left[\ell_\alpha (Y - G(\alpha))\right] - \frac{1}{\alpha}\E [\ell_\alpha (Y - g(\alpha))] \right) d\alpha \nonumber  \\
&\leq& 0,
\eean
then $0 \leq \mathcal{R}(G, G) \leq \mathcal{R}(g, G)$. 

Now we will show that $\mathcal{R}(g, G) < \infty$. 
Assumption A and the assumption of proposition 5 imply that
\bean
|F(g(\alpha)) g(\alpha)| \asymp |\alpha^{-q}|^{-r} (\alpha^{-q}) = \alpha^{q(r-1)}
\eean
for some $r>1$ ane $0<q<1$ as $\alpha \downarrow 0$. Because $|F(g(\alpha)) g(\alpha)| \asymp \alpha^{q(r-1)} $ with $q(r-1)>0$ as  $\alpha \downarrow 0$ and $\int_0^1 |g(\alpha)|d\alpha <\infty$,
\bea \label{proof:lemma1_1}
\left | \int_0^1 \frac{1}{\alpha}F(g(\alpha))g(\alpha)d\alpha \right| < \infty.
\eea
Similarly, it is shown that $\int_{-\infty}^{g(\alpha)} y dF(y) \asymp \alpha^{q(r-1)}$ as $\alpha \downarrow 0$, which implies
\bea \label{proof:lemma1_2}
\left| \int_0^1 \frac{1}{\alpha}  \int_{-\infty}^{g(\alpha)} y dF(y) \right|  < \infty
\eea
with $\E(Y)<\infty$. Then,
\bean
&& \mathcal{R}(g,G) \\ &=& \int_0^1 \frac{1}{\alpha} \E\left[\ell_\alpha (Y - g(\alpha))\right] d\alpha \\ 
&=& \int_0^1 \frac{1}{\alpha} \left[\int_{-\infty}^\infty (\alpha - I(y \leq g(\alpha )))(y - g(\alpha)) dF(y) \right] d\alpha \\ 
&=& \int_0^1 \frac{1}{\alpha} \left[\int_{-\infty}^\infty \alpha y - \alpha g(\alpha) + g(\alpha) I(y \leq g(\alpha )) - yI(y \leq g(\alpha )) dF(y) \right] d\alpha \\ 
&=& \int_0^1 \frac{1}{\alpha}\left[ \alpha \E(Y) - \alpha g(\alpha) + g(\alpha) F(g(\alpha)) - \int_{-\infty}^{g(\alpha)} y dF(y) \right] d\alpha \\
&=& \E(Y) - \int_0^1 g(\alpha)d\alpha  + \int_0^1 \frac{1}{\alpha} F(g(\alpha))g(\alpha)d\alpha 
- \int_0^1 \frac{1}{\alpha} \int_{-\infty}^{g(\alpha)} y dF(y) d\alpha \\
&<& \infty.
\eean
The last equality holds because $\E|Y|<\infty$, $\int_0^1 |g(\alpha)| d\alpha <\infty$, \eqref{proof:lemma1_1}, and \eqref{proof:lemma1_2}.

\end{proof}
\subsection{Proof of Proposition 6}
\begin{proof}
By Proposition 5, $R(G,G)<\infty$  and 
\bea \label{def: D}
&&\mathcal{R}(G,G) - \mathcal{R}(g, G) \nonumber \\
&=&\int_0^1 \frac{1}{\alpha} \E\left[\ell_\alpha (Y - G(\alpha))\right] - \frac{1}{\alpha}\E [\ell_\alpha (Y - g(\alpha))] d\alpha \nonumber  \\
&=& \int_0^1  \bigg[ g (\alpha) - G(\alpha) + \frac{1}{\alpha}\int y(I(y \leq g(\alpha)) - I(y \leq  G(\alpha))) \nonumber \\ 
&& ~~~~~~~~~~~~~~~~~~~~~~~~~~~~~~~~~ + I(y \leq G(\alpha))G(\alpha) - I(y \leq g(\alpha))g(\alpha) dF(y) \bigg] d\alpha.
\eea
Let the integrand of \eqref{def: D} be $D(\alpha; g, G)$, and then it suffices to prove that 
$\int_0^1 D(\alpha; g, G)d\alpha = -\infty$
for an arbitrary continuous, bounded, and non-decreasing function $g_\theta$. \\
Define a function $m: (0, 1) \mapsto \mathbb{R}$ satisfying
\[
\min(g(\alpha), G(\alpha)) \leq m(\alpha) \leq \max(g(\alpha), G(\alpha)).
\] 
Then, it follows that 
\bea
&& \int y\big(I(y \leq g(\alpha)) - I(y \leq G(\alpha)) \big) dF(y)  \label{eq: ori-integral} \\ 
&\leq& \int m(\alpha)\big(I(y \leq g(\alpha)) - I(y\leq G(\alpha))\big) dF(y). \label{eq: upper-integral} 
\eea
By replacing the term \eqref{eq: ori-integral} in $D(\alpha; g, G)$ by \eqref{eq: upper-integral}, we obtain for all $\alpha \in (0, 1)$
\bean \label{eq: upperbound}
&& D(\alpha; g, G)  \\  
 &\leq& g (\alpha) - G(\alpha) + \frac{1}{\alpha}\int \bigg[ m(\alpha) (I( y\leq g(\alpha)) - I(y \leq  G(\alpha))) \nonumber \\ 
&+& I(y \leq G(\alpha))G(\alpha) - I(y \leq g(\alpha))g(\alpha)\bigg] dF(y) \nonumber  \\
&=&  g(\alpha) - G(\alpha) +  \frac{1}{\alpha}\bigg[m(\alpha)(F(g(\alpha)) - \alpha) + \alpha G(\alpha) - F(g(\alpha))g(\alpha) \bigg] \nonumber \\
&=& (g(\alpha) - m(\alpha))\left(1 - \frac{F(g(\alpha))}{\alpha} \right) = D^*(\alpha; g, G).
\eean

Note that $g:(0,1) \mapsto \mathbb{R}$ is a continuous, bounded, and non-decreasing function, and we can set $x_1 \in \mathbb{R}$  and $\alpha_1\in (0,1)$ such that $g(\alpha) = x_1$ for all $0<\alpha \leq \alpha_1$. While, $G(\alpha)$ is unbounded, non-decreasing  function, we can set $\alpha_1'$ such that  $G(\alpha)\leq x_1-\epsilon$ for $0<\alpha\leq \alpha_1'$.  Let $\alpha_L = \min(\alpha_1, \alpha_1')$. Then,  by definition of $m(\alpha)$, we can choose the $m(\cdot)$ satisfying $m(\alpha) = x_1 -\epsilon/2$ for $0<\alpha \leq \alpha_L$, which implies  $g(\alpha) - m(\alpha) = \epsilon/2$
for $\alpha \in (0, \alpha_L)$. Similarly, we can set $\alpha_U\in (\alpha_L,1)$, such that  $g(\alpha) - m(\alpha) = -\epsilon/2$. Because $g(\cdot)$ and $G(\cdot)$ is continuous, $g(\alpha)-m(\alpha)$ is bounded on $[\alpha_L, \alpha_U]$. Thus, without loss of generality, we can assume that $g(\alpha) - m(\alpha)$ is bounded.  

Here, it is shown that $
\int_0^{\alpha_L} D^*(\alpha;g, G)d\alpha = -\infty$, 
since $\underset{\alpha \downarrow 0}{\lim} ~ F(g(\alpha)) = F(x_1) >0$.  Because  
$\int_{\alpha_L}^1 D^*(\alpha; g, G)d\alpha < \infty$, we can conclude that 
$\int_0^1 D^*(\alpha; g, G)d\alpha = -\infty$, which completes the proof.
\end{proof}

\subsection{Proof of Proposition 7}
\begin{proof}
First, we induce a lower bound of $D(\alpha; g, G)$. Consider two cases as follows:
\ben
\item $g(\alpha) \geq G(\alpha)$ 
\item $G(\alpha) > g(\alpha)$
\een
In both cases, we obtain 
\bea
&& \int G(\alpha)(I(y \leq g(\alpha)) - I(y \leq G(\alpha))) dF(y) \label{eq: lower-integral} \\ 
&\leq& \int y(I(y \leq g(\alpha)) - I(y \leq G(\alpha))) dF(y)  \nonumber \\ 
&\leq& \int g(\alpha)(I(y \leq g(\alpha)) - I(y\leq G(\alpha))) dF(y). \label{eq: for proper}
\eea
By replacing \eqref{eq: ori-integral} with \eqref{eq: lower-integral} in $D(\alpha; g, G)$ and using $D^*(\alpha; g, G)$, we obtain
\bean
\int_\eta^1 (g(\alpha) - G(\alpha))\left(1 - \frac{F(g(\alpha))}{\alpha} \right) d\alpha \leq \int_\eta^1  D(\alpha; g, G) d\alpha 
\leq \int_\eta^1  D^*(\alpha; g, G ) d\alpha.
\eean
There is a constant $K$ such that $0 < F(g(\alpha))/\alpha \leq K < \infty$ where $\alpha \in [\eta, 1)$. Under Assumption B, 
\[
\int_\eta^1 \left|g(\alpha)\frac{F(g(\alpha))}{\alpha}\right| d\alpha \leq K \int_\eta^1 |g(\alpha)| d\alpha < \infty.
\]
Under $\E|Y| <\infty$ implying $\int_0^1 G(\alpha) d\alpha < \infty$, we have $\int_\eta^1 |G(\alpha)| d\alpha < \infty$. Then,
\[
\int_\eta^1 \left| G(\alpha) \frac{F(g(\alpha))}{\alpha}\right| d\alpha  < \infty. 
\]  
Therefore, there are constants, $-\infty < M_1 \leq M_2 \leq 0$ such that 
\bean
M_1 = \int_\eta^1 (g(\alpha) - G(\alpha))\left(1 - \frac{F(g(\alpha))}{\alpha} \right) d\alpha \leq
\int_\eta^1  D(\alpha; g, G) d\alpha \leq \int_\eta^1  D^*(\alpha; g, G ) d\alpha = M_2.
\eean
Thus, 
\[
\left| \int_\eta^1 D(\alpha; g, G) \right | < \infty,
\]
which concludes $\mathcal{R}_\eta(g, G) < \infty$.

Because  replacing the term \eqref{eq: ori-integral} in $D(\alpha; g, G )$ with \eqref{eq: for proper} leads to  $\mathcal{R}_\eta(G,G) - \mathcal{R}_\eta(g, G) \leq 0$, we obtain $0 \leq \mathcal{R}_\eta(G,G) \leq \mathcal{R}_\eta(g, G) < \infty$.
Finally, $\underset{\eta \downarrow 0}{\lim} ~ \mathcal{R}_\eta(G, G) = \mathcal{R}(G, G)$ by $\mathcal{R}(G, G) < \infty$.
\end{proof}

\subsection{Proof of Theorem 1} 
\begin{proof}
By Proposition 7, $\mathcal{R}_\eta(g_{\theta}, G) < \infty$.
In addition, $\mathcal{R}_\eta(g_{\theta}, G) = \int_\eta^1 \alpha^{-1} \E [\ell_{\alpha}  (Y - g_{\theta}(\alpha))]d\alpha = \E \left(\int_\eta^1 \alpha^{-1} \ell_{\alpha}(Y - g_{\theta}(\alpha))d\alpha \right)$ by Fubini-Tonelli theorem, such that $E|\int_\eta^1 \alpha^{-1} \ell_{\alpha}(Y -g_{\theta}(\alpha))d\alpha|<\infty$. Thus, it follows that 
\bean
\frac{1}{n}\sum_{i=1}^n \int_\eta^1 \alpha^{-1} \ell_{\alpha}(y_i -g_{\theta}(\alpha))d\alpha \rightarrow  E\left( \int_\eta^1 \alpha^{-1} \ell_{\alpha}(Y -g_{\theta}(\alpha))d\alpha \right)
\eean
almost surely as $n \rightarrow \infty$ by the law of large numbers.
\end{proof}

\subsubsection*{S1.8 Proof of Theorem 2}
\begin{proof}
Let $S_\theta(y; \eta) = \int_\eta^1 \alpha^{-1} \ell_\alpha(y - g_{\theta}(\alpha)) d\alpha$. Suppose that the following conditions hold: 
\ben
    \item For each $\theta\in \Theta$ there exist  $\omega:[\eta,1) \rightarrow \mathbb{R}$, and a small neighborhood $U$ of $\theta$ such that $\underset{\theta \in U}{\sup}\frac{1}{\alpha}|g_{\theta}(\alpha)| \leq  \omega(\alpha)$ and $\int_\eta^1 | \omega(\alpha)| d\alpha <\infty$.
    \item For a sequence $\theta_n$ going to $\theta$, a pointwise limit $ g_{\theta}(\alpha) = \underset{\theta_n \rightarrow \theta}{\lim} g_{\theta_n}(\alpha)$ exists. 
\een
\bean
&& |S_{\theta_n}(y; \eta) - S_\theta(y;\eta)| \\  
&=& \Big|\int_\eta^1 \alpha^{-1} \left( \ell_\alpha(y - g_{\theta_n}(\alpha)) - \ell_\alpha(y - g_{\theta}(\alpha))  \right) d\alpha \Big| \\ 
&=& \Big|\int_\eta^1 \alpha^{-1}\Big[- (\alpha - I(y \leq g_{\theta_n}(\alpha)))(g_{\theta_n}(\alpha) - g_{\theta}(\alpha)) \\ &+&   \int_0^{G_{\theta_n}(\alpha) - G_{\theta}(\alpha)} I(y \leq g_{\theta_n}(\alpha) + s) - I(y \leq g_{\theta_n}(\alpha)) ds \Big] d\alpha \Big|\\
&\leq& 2 \int_\eta^1 \alpha^{-1} |(g_{\theta_n}(\alpha) - g_{\theta}(\alpha)) | d\alpha ,
\eean
where the second equality follows from  Knight's identity. \\
Because $\alpha^{-1} |g_{\theta_n}(\alpha)- g_{\theta}(\alpha)|$ is bounded from above by $2 \omega(\alpha)$,  
\bean
\underset{\theta_n \rightarrow \theta}{\lim} \int_\eta^1 \alpha^{-1} |(g_{\theta_n}(\alpha) - g_{\theta}(\alpha)) | d\alpha = \int_\eta^1 \alpha^{-1} | \underset{\theta_n \rightarrow \theta}{\lim} (g_{\theta_n}(\alpha) - g_{\theta}(\alpha)) | d\alpha = 0 
\eean
by dominated convergence theorem. Thus,  $S_\eta(\theta, y)$ is continuous at $\theta$. Then, by Theorem 5.14 \citep{van2000asymptotic}, 
$$\Pr\left(d(\hat \theta_n, \Theta_0) >\epsilon, \hat \theta_n \in \Theta \right) \rightarrow 0 $$
as $ n \rightarrow \infty$.
\end{proof}
\subsection{Proof of Proposition 8}
\begin{proof}
See Corollary 1 \citep{ruszczynski2006optimization}.
\end{proof}

\subsection{Proof of Theorem 3}
\begin{proof}
We introduce a beta distribution as $\varphi = b(\alpha; s,h) \sim \text{Beta}(s, h)$. Suppose that $\alpha \sim b(\alpha; s,h), ~ s \geq 1$ and $h=1$ then, the pessimistic risk measure $\varrho_{(s,h)}(Y)$ can be written as follows:
\bean
\varrho_{(s,1)}(Y) &=& - \int_0^1  \frac{b(\alpha; s,1)}{\alpha} \int_0^\alpha G(t) dt d\alpha \\
&=& - \int_0^1  \frac{1}{B(1, s, 1)} \alpha^{s-2} \int_{-\infty}^{G(\alpha)} y f(y) dy d\alpha \\
&=& - s \int_0^1 \int_{-\infty}^{t} F(t)^{s-2} y f(y) f(t) dy dt \\
&=& - s \int_{-\infty}^{\infty} y f(y) \int_{y}^{\infty} F(t)^{s-2}f(t) dt dy \\
&=& - \frac{s}{s-1} \int_{-\infty}^{\infty} y f(y) (1-F(y)^{s-1}) dy, 
\eean
where $s \neq 1$.\\
Under Assumption A, $\int_{-\infty}^{\infty} | y f(y) (1-F(y)^{s-1})| dy < \infty$. Therefore, the exchange of double integrals leads to 

\bean
\varrho_{(s,1)}(Y) &=& - \int_0^1 \frac{1}{B(1, s, 1)}\alpha^{s-2}\int_0^\alpha G(t)dt d\alpha \\
&=&  -\int_0^1 s \alpha^{s-2} \int_0^\alpha G(t) dt d\alpha\\
&=& - \int_0^1 G(t) s \int_t^1 \alpha^{s-2} d\alpha dt  \\
&=& - \int_0^1 G(t) s \left[\frac{\alpha^{s-1}}{s-1}\right]_{\alpha=t}^1 dt \\ 
&=& - \int_0^1 \frac{s}{s-1}G(t)(1-t^{s-1}) dt.
\eean
Let $G(t) = y$, then 
\bea 
&& - \int_0^1 \frac{s}{s-1}G(t)(1-t^{s-1}) dt \nonumber \\ 
&=& -\int_{-\infty}^\infty \frac{s}{s-1}  y(1-\{F(y)\}^{s-1}) f(y) dy \nonumber\\ 
&=&  -\int_{-\infty}^\infty \frac{s}{s-1} (y f(y) - y\{F(y)\}^{s-1}f(y))dy \nonumber \\ 
&=& - \frac{s}{s-1}\left[\E(Y) - \E(Y\{F(Y)\}^{s-1})\right] \label{eq:beta_upr}.
\eea
The cumulative distribution function and probability density function of the GEV distribution are given as follows: 
\bean
F(y; \xi, \zeta, \kappa) &=& \exp \left( - \left\{1 - \kappa \left(\frac{y - \xi}{\zeta} \right) \right\}^{1/
\kappa} \right), \\
f(y; \xi, \zeta, \kappa) &=& \exp \left( - \left\{1 - \kappa \left(\frac{y - \xi}{\zeta} \right)\right\}^{1/
\kappa} \right) \frac{1}{\zeta} \left\{ 1-\kappa \left(\frac{x-\xi}{\zeta} \right) \right\}^{1/\kappa - 1} \\ 
\text{s.t.} && \kappa \neq 0 ~ \text{and} ~ 1- \kappa\left(\frac{y -\xi}{\zeta}\right) > 0,
\eean
where $\xi, \zeta,$ and $\kappa$ are the location, scale, and shape parameters, respectively.

For $Y \sim \text{GEV}(\xi, \zeta, \kappa)$ with $\kappa > -1$, the probability-weighted moments \citep{jeon2011expected}, $\E(Y\{F(Y)\}^{s-1})$ is written in a closed form as
\bean
\E(Y\{F(Y)\}^{s-1}) = \frac{1}{s}[\xi + \zeta(1 - s^{-\kappa}\Gamma(1+\kappa))/\kappa],
\eean
and expectation of $Y$ is $\E(Y) = \xi + \frac{\zeta}{\kappa}(1-\Gamma(1+\kappa))$.\\
Then plugging them into above equation \eqref{eq:beta_upr},
\bea
&& \varrho_{(s,1)}(Y) \nonumber \\ 
&=& -\frac{s}{s-1}\left[\xi +\frac{\zeta}{\kappa}(1-\Gamma(1+\kappa) - \frac{1}{s}\{\xi + \zeta(1-s^{-\kappa}\Gamma(1+\kappa))/\kappa\}  \right] \nonumber \\
&=& \frac{\zeta\Gamma(1+\kappa)}{\kappa} \cdot \frac{(s - s^{-\kappa})}{s-1} -\xi -\frac{\zeta}{\kappa} = \frac{a}{\kappa} \cdot \frac{(s - s^{-\kappa})}{s-1} + c \label{eq: varrho-s}
\eea
where $a>0$ and $c$ are constants which do not depend on $s$.\\
Then, for $s > 1$,  
\bean
\left(\frac{s - s^{-\kappa}}{s-1}\right)' = \frac{(\kappa+1)s^{-\kappa} - \kappa s^{-\kappa -1} - 1}{(s-1)^2} = \begin{cases} > 0,  & -1 < \kappa < 0, \\ 
< 0, & \kappa > 0.
\end{cases}
\eean

Now we show that $\varrho_{(s, 1)}(Y)$ is right-continuous at $s=1$. Let $(1 - \kappa ((y - \xi)/\zeta))^{1/\kappa} = x$ then, 
\bean
\varrho_{(1, 1)}(Y) &=& \int y (-\log F(y)) f(y) dy \\
&=& \int y \left\{ 1 - \kappa \left(\frac{y - \xi}{\zeta} \right)\right\}^{1/\kappa} \exp \left( - \left\{1 - \kappa \left(\frac{y - \xi}{\zeta} \right)\right\}^{1/
\kappa} \right)  \\ 
&\times& \frac{1}{\zeta} \left\{ 1-\kappa \left(\frac{x-\xi}{\zeta} \right) \right\}^{1/\kappa - 1}dy \\ 
&=& \int_0^\infty \zeta \left(\frac{1-x^\kappa}{\kappa} + \xi \right) (-x) \exp(-x) dx = \frac{\zeta \Gamma(2+\kappa)}{\kappa} - \xi - \frac{\zeta}{\kappa},
\eean
and the limit of \eqref{eq: varrho-s} as $s \downarrow 1$ is given by,
\bean
\lim_{s \downarrow 1} \varrho_{(s, 1)}(Y) &=& \lim_{s \downarrow 1}\frac{s - s^{-\kappa}}{s - 1}\frac{\zeta \Gamma(1+\kappa)}{\kappa}   - \xi - \frac{\zeta}{\kappa} \\
&=& (\kappa + 1)\frac{\zeta \Gamma(1+\kappa)}{\kappa} - \xi - \frac{\zeta}{\kappa} \\ 
&=& \varrho_{(1, 1)}(Y).
\eean
Therefore, $\varrho_{(s, 1)}$ is right-continuous at $s=1$ when $\kappa \neq 0$. Similarly, we can show that $\varrho_{(s, 1)}$ is decreasing w.r.t $s > 1$ and right-continuous at $s = 1$ when $\kappa=0$, i.e., $Y$ is a Gumbel random variable. As a consequence, $\varrho_{(s,1)}(Y)$ is a monotonically decreasing function of $s$ where $s \geq 1$.
\end{proof}

\subsection{Proof of Proposition 9}
\begin{proof}
We have
\bean
&& \int_0^1 \frac{1}{\alpha}\E(\ell_\alpha (Y - g(\alpha)))b(\alpha; s, h) d\alpha \\
&=& \int_0^1 \frac{b(\alpha; s,h)}{\alpha} \bigg [\int \alpha y - \alpha g(\alpha) + I(y \leq g(\alpha))g(\alpha) - I(y\leq g(\alpha)) y dF(y) \bigg] d\alpha \\
&=& \E(Y) - \int_0^1 b(\alpha; s, h) g(\alpha) d\alpha + \int_0^1 \frac{b(\alpha; s,h)}{\alpha}\left(F(g(\alpha))g(\alpha) - \int_{-\infty}^{g(\alpha)} y dF(y) \right) d\alpha.
\eean
When $s \geq 1$ and $h \geq 1$, then $ \sup_{\alpha\in (0,1)}b(\alpha; s, h) < \infty$. Because $\int_0^1 |g(\alpha)| d\alpha < \infty$,
\[
\int_0^1 |b(\alpha; s, h) g(\alpha) |d\alpha < \infty.
\]
Now we prove that 
\bean
\left | \int_0^1 \frac{b(\alpha; s,h)}{\alpha}F(g(\alpha))g(\alpha)d\alpha \right| &<& \infty \\
\left | \int_0^1 \frac{b(\alpha; s,h)}{\alpha} \int_{-\infty}^{g(\alpha)} y dF(y)  d\alpha \right | &<& \infty.
\eean
There are two possible cases listed as follows:
\ben
    \item $g$ is not bounded below.
    \item $g$ is bounded below.
\een
If $g$ is not bounded below, $ b(\alpha; s, h)\int_{-\infty}^{g(\alpha)} ydF(y) \asymp \alpha^{(s-1)+q(r-1)}$ as $\alpha \downarrow 0$ for $s\geq 1$, $0<q<1$, and $r>1$, which implies
\bean 
 \left | \int_0^1 \frac{b(\alpha; s,h)}{\alpha} \int_{-\infty}^{g(\alpha)} y dF(y)  d\alpha \right | < \infty.
\eean
Similarly, 
\bea \label{cond: finite2}
\left | \int_0^1 \frac{b(\alpha; s,h)}{\alpha}F(g(\alpha))g(\alpha)d\alpha \right| < \infty,
\eea
because $b(\alpha; s,h)F(g(\alpha))g(\alpha)\asymp \alpha^{(s-1)+q(r-1)}$ 
as $\alpha \downarrow 0$ for $s\geq 1$, $0<q<1$.
If $g$ is bounded below, let $\underset{\alpha \downarrow 0}{\lim} ~ g(\alpha) = k$. Assume that $s > 1$, then
\[
b(\alpha; s,h)\int_{-\infty}^{g(\alpha)} y dF(y) \asymp \alpha^{s-1},~~\mbox{as } \alpha \downarrow 0,
\]
which also implies \eqref{cond: finite2}. Therefore, for $s>1$,
\[
 \int_0^1 \frac{1}{\alpha}\E(\ell_\alpha (Y - g(\alpha)))b(\alpha; s, h) d\alpha  < \infty.
\]
\end{proof}

\subsection{Proof of Theorem 4}

\begin{lemma} \label{lemma1}
Let $\bbeta \in \{\bbeta \in \mathbb{R}^p : |\bbeta|_\infty \leq \lambda^{*}, \lambda^* > 0 \}$ and $\bx$ be $p$-dimensional vector of $\sigma^2$-sub Gaussian random variables. Then, 
\bean
\mathbb{E}\left[\sup_{\left|\bbeta\right|_{\infty} \leq \lambda^*} \left|\bx^\top \bbeta \right| \right] &\leq& 2\sqrt{p}\lambda^* \sigma \sqrt{2p\log\left(1+\frac{4}{\sqrt{p}\lambda^*}\right)}, \\
\mathbb{E}\left[\sup_{\left|\bbeta\right|_{\infty} \leq \lambda^*} \max_{ j=1, \dots, p} \left|\bx^\top \bbeta \right| |\bx_{j}| \right] &\leq& 7\lambda^{*}p^{3/2}\sigma^2 \sqrt{\log\left(1+\frac{4}{\sqrt{p}\lambda^*}\right)},
\eean
where $\bx_j$ is the $j$-th element of $\bx$.

\end{lemma}

\begin{proof}
Let $B_r(x)$ be an Euclidean ball of radius $r$ and center $x$. For notation simplicity, $B_{r}(x)$ when $x$ is the origin is denoted by $B_r$. 
\bean
\sup_{\left|\bbeta\right|_{\infty} \leq \lambda^*}\left|\bx^\top \bbeta\right| \leq \sup_{\bbeta \in B_{\sqrt{p}\lambda^*}} \left|\bx^\top \bbeta\right| = \sup_{\bbeta \in B_{\sqrt{p}\lambda^*}} \bx^\top \bbeta.
\eean
Consider a covering of $B_{\sqrt{p}\lambda^*}$ with at most $\lceil (1+4/(\sqrt{p}\lambda^*))^p\rceil = C$ balls of radius $\sqrt{p}\lambda^*/2$, denoted by $\bbeta^{(1)}, \dots, \bbeta^{(C)}$. Then, we have 
\bea
\sup_{\bbeta \in B_{\sqrt{p}\lambda^*}} \bx^\top \bbeta \leq \max_{c = 1, \dots, C} \bx^\top \bbeta^{(c)} + \sup_{\bbeta \in B_{\sqrt{p}\lambda^*/2}} \bx^\top \bbeta. \label{eq:subg_split}
\eea
Under the condition $\bbeta \in B_{\sqrt{p}\lambda^*/2}$, we have $\bbeta^\top \bx \sim \text{subG}(p\lambda^{*2} \sigma^2/4)$. Then, 
\bean
\mathbb{E} \left[ \sup_{\bbeta \in B_{\sqrt{p}\lambda^*}} 
\bx^\top \bbeta \right] &\leq& 2 \mathbb{E} \left[\max_{c = 1, \dots, C} \bx^\top \bbeta^{(c)}\right] \\ 
&\leq& \frac{\sqrt{p} \lambda^* \sigma}{2}\sqrt{2p\log\left(1+\frac{4}{\sqrt{p}\lambda^*}\right)}.
\eean
By Cauchy-Schwarz Inequality, we have
\bea
&& \E \left[\sup_{\left|\bbeta\right|_{\infty} \leq \lambda^*}  \left| \bx^\top \bbeta \right| \max_{j=1, \dots, p}|\bx_{j}|\right] \nonumber \leq \left(\E\left[\sup_{\left|\bbeta\right|_{\infty} \leq \lambda^*} \left| \bx^\top \bbeta \right|^2 \right]\right)^{1/2} \left(\E\left[\max_{j=1, \dots, p} |\bx_{j}|^2\right] \right)^{1/2}. \label{eq:cauchy_schwarz}
\eea
\bea
\E\left[\max_{j=1, \dots, p} |\bx_{j}|^2\right]  &=& 8\sigma^2\E\left[ \log\left( \exp(\max_{j=1, \dots, p}  \bx_{j}^2 /8\sigma^2  )\right) \right] \nonumber \\ 
&\leq& 8\sigma^2 \log \left(\E\left[ \exp( \max_{j=1, \dots, p} \bx_{j}^2/8\sigma^2)\right]\right) \nonumber \\ 
&\leq& 8\sigma^2 \log \left(\E\left[\sum_{j=1}^p \exp(\bx_{j}^2/8\sigma^2)\right]\right) \nonumber \\
&\leq& 8\sigma^2 \log(2p). \label{eq:squared_ineq1}
\eea
From the inequality \eqref{eq:subg_split}, 
\bean
\sup_{\bbeta \in B_{\sqrt{p}\lambda^*}} \left|\bx^\top \bbeta\right|^2 &\leq& \max_{c = 1, \dots, C} \left|\bx^\top \bbeta^{(c)}\right|^2 + \sup_{\bbeta \in B_{\sqrt{p}\lambda^*/2}} \left|\bx^\top \bbeta\right|^2 \\
&=& \max_{c = 1, \dots, C} \left|\bx^\top \bbeta^{(c)}\right|^2 + \frac{1}{4}\sup_{\bbeta \in B_{\sqrt{p}\lambda^*}} \left|\bx^\top \bbeta\right|^2.
\eean
Thus, we have
\bea
&& \E\left[\sup_{\left|\bbeta\right|_{\infty} \leq \lambda^*} \left| \bx^\top \bbeta \right|^2 \right] \nonumber \\
&\leq& \frac{4}{3}  \E\left[ \max_{c=1,\dots, C} |\bx^\top \bbeta^{(c)}|^2 \right]  \nonumber \\
&\leq&  \frac{128}{27}p\lambda^{*2}\sigma^2 \log\left(\E\left[\exp \max_{c=1,\dots, C} 9\left|\bx^\top \bbeta \right|^2 /(32\lambda^{*2}p\sigma^2) \right]\right) \nonumber \\ 
&\leq&  \frac{128}{27}p\lambda^{*2}\sigma^2 \log\left(\E\left[\exp \sum_{c=1}^C 9\left|\bx^\top \bbeta \right|^2 /(32\lambda^{*2}p\sigma^2) \right]\right) \nonumber\\ 
&\leq& \frac{128}{27} p\lambda^{*2}\sigma^2\sqrt{\log(2C)}. \label{eq:squared_ineq2}
\eea
The final inequalities in the derivations of \eqref{eq:squared_ineq1} and \eqref{eq:squared_ineq2} are derived from the following inequality: 
\[
\E\left[X/(8\sigma^2)\right] \leq 2,
\]
where $X \sim \mbox{subG}(\sigma^2)$.\\
Plugging \eqref{eq:squared_ineq1} and \eqref{eq:squared_ineq2} into \eqref{eq:cauchy_schwarz}, we have 
\bean
\E \left[\sup_{\left|\bbeta\right|_{\infty} \leq \lambda^*}  \left| \bx^\top \bbeta \right| \max_{j=1, \dots, p}|\bx_{j}|\right] \leq 7  \lambda^* p^{3/2} \sigma^2\sqrt{\log \left(1+\frac{4}{\sqrt{p}\lambda^*}\right)}.
\eean

\end{proof}
\begin{definition} Let $(\Psi, \|\cdot\|)$ be a metric space. For $\epsilon > 0$, $\{\psi^i\}_{i=1}^N$ is called $\epsilon$-cover (or $\epsilon$-net) of $\Psi$ if
\[
 \forall \psi \in \Psi, ~ \exists i ~\mbox{s.t.}~ \| \psi^i - \psi\| \leq \epsilon.
\]
The smallest size of $\{\psi^i\}_{i=1}^N$ is the $\epsilon$-covering number, and we denote it by $N(\Psi, \|\cdot\|, \epsilon)$.
\end{definition}
\begin{lemma}[19.7 \citep{van2000asymptotic}] \label{lemma:2} Let $\mathcal{S} \in \{S_\psi 
 : \psi \in \Psi \}$ be a function class with a compact subset $\Psi$. For every $\psi_1, \psi_2 \in \Psi$, assume that there exists a function $c(x)$ with $\E[|c(x)|] < \infty$ such that 
 \[
|S_{\psi_1}(x) - S_{\psi_2}(x)| \leq |c(x)|\|\psi_1 - \psi_2\|.
 \]
Then, $N(\mathcal{S}, |\cdot|, \E[|c(x)|]\epsilon) \leq N(\Psi, \|\cdot\|, \epsilon/2)$.
\end{lemma}
\begin{proof}
Because $\Psi$ is compact set, let $\{\psi^i\}_{i=1}^N$ be a $\epsilon/2$-cover of $\Psi$. Define $S^u_i$ and $S^l_i$ as upper and lower bounds of $S_{\psi^i}$ as follows:
\bean
S^u_i(x) &=& S_{\psi^i}(x) + \frac{1}{2}|c(x)|, \\
S^l_i(x) &=& S_{\psi^i}(x) - \frac{1}{2}|c(x)|.
\eean
By the assumptions of $\{\psi^i\}_{i=1}^N$ and $\mathcal{S}$, we have 
\bean
|S_\psi(x) - S_{\psi^i}(x)| \leq |c(x)| \|\psi - \psi^i\| \leq \frac{\epsilon}{2}|c(x)|. 
\eean
Therefore, $N(\mathcal{S}, |\cdot|, \E[|c(x)|]\epsilon) \leq N(\Psi, \|\cdot\|, \epsilon/2)$.
\end{proof}

\noindent \textbf{Proof of Theorem 4}
\begin{proof}
 Let $S_{\psi}(\bx; \eta) = \int_\eta^1 \alpha^{-1} \ell_\alpha\left(\bx^\top \bbeta - g_{\theta}(\alpha)\right) d\alpha$ and $\tilde{\alpha} =  g^{-1}_\theta(\bx^{\top}\bbeta) = \frac{\bx^{\top}\bbeta - \gamma + \sum_{m=0}^{\tilde{m}} b_m d_m}{\sum_{m=0}^{\tilde{m}} b_m}$ with $d_{\tilde{m}} \leq \tilde{\alpha} < d_{\tilde{m}+1}$, and $\tilde{m}$ is the largest knot position index such that $g_\theta(d_{\tilde{m}}) \leq g_\theta(\tilde{\alpha})$. Then, we derive the closed form of $S_\psi(\bx ; \eta)$ as follows:
\bean
&& S_\psi(\bx;\eta)  \\
&=& \int_\eta^1 \alpha^{-1} \ell_\alpha(\bx^\top \bbeta - g_{\theta}(\alpha)) d\alpha   \\
&=& \int_\eta^1 \frac{1}{\alpha} \ell_\alpha \left(\bx^{\top} \bbeta -  \gamma - \sum_{m=0}^M b_m(\alpha - d_m)_+\right) d\alpha   \\
&=& \int_\eta^{\tilde{\alpha}} \left(\bx^{\top} \bbeta - \gamma - \sum_{m=0}^M b_m(\alpha - d_m)_+\right) d\alpha + 
\int_{\tilde{\alpha}}^1 \left(1-\frac{1}{\alpha}\right) \left(\bx^{\top} \bbeta - \gamma - \sum_{m=0}^M b_m(\alpha - d_m)_+ \right) d\alpha  \\
&=& \int_\eta^1 \left(\bx^{\top} \bbeta- \gamma - \sum_{m=0}^M b_m(\alpha - d_m)_+\right) d\alpha - \int_{\tilde{\alpha}}^1 \frac{1}{\alpha}\left(\bx^{\top} \bbeta- \gamma - \sum_{m=0}^M b_m(\alpha - d_m)_+ \right) d\alpha   \\
&=& (1-\eta +\log\tilde{\alpha})(\bx^{\top}\bbeta-\gamma) + \sum_{m=0}^M b_m\left(1-\frac{(1-d_m)^2}{2} - \max(\tilde{\alpha}, d_m) + \max(\log \tilde{\alpha}, \log d_m)d_m\right). 
\eean
Let $\sum_{m=0}^{\tilde{m}} b_m = \bar{b}$, and $\sum_{m=0}^{\tilde{m}} b_m d_m = \bar{d}$. By assumptions, $\tilde{\alpha} \geq \eta \in (0, 1)$, and $\bar{b} > 0$. Then, the $l_1$ norms of derivatives of $S_\psi$ and their upper bounds are as follows:
\bean 
\left|\frac{\partial S_\psi(\bx; \eta)}{\partial \gamma} \right| &=&  \left |- \left(\frac{\bx^\top \bbeta - \gamma + \bar{b}}{\tilde{\alpha} \bar{b}} - \eta + \log \tilde{\alpha} \right)\right| \\ &\leq& \left|\frac{\bx^\top \bbeta}{\eta\tilde{b}}\right| + \left(\frac{|\gamma| + \hat{b} }{\eta \tilde{b}} + \eta - \log \eta \right) = C_1(\bx, \bbeta), \\ 
\left|\frac{\partial S_\psi(\bx;\eta)}{\partial \bbeta}\right| &=& \left| \left(\frac{\bx^\top \bbeta - \gamma + \bar{b}}{\tilde{\alpha} \bar{b}} - \eta + \log \tilde{\alpha}\right)\bx \right| \leq C_1(\bx, \bbeta)|\bx| = C_2(\bx, \bbeta), \\
\left|\frac{\partial S_\psi(\bx; \eta)}{\partial b_m}\right| &=& \begin{cases}
 \bigg| 1-\frac{(1-d_m)^2}{2} - d_m + d_m\log d_m \bigg|,~  \text{if} ~ m > \tilde{m},\\
\bigg|1 - \frac{(1-d_m)^2}{2} - \tilde{\alpha} + d_m \log \tilde{\alpha} +\left(\frac{\bar{b} d_m - \bar{d}}{\bar{b}}\right)\left(1 + \frac{\bx^\top \bbeta - \gamma - d_m \bar{b}}{\tilde{\alpha} \bar{b}} \right) \bigg|, \text{if o.w.},
\end{cases} \\
&\leq& 1 - \log \eta + \frac{\hat{b}}{\tilde{b}} + \frac{\hat{b}}{\tilde{b}} \left(\left|\frac{\bx^\top \bbeta}{\eta\tilde{b}}\right| + \frac{|\gamma| + \hat{b}}{\eta\tilde{b}} \right) = C_3(\bx, \bbeta),
\eean
where $\hat{b} = \underset{\tilde{m}}{\max} \sum_{m=0}^{\tilde{m}} b_m$ and $\tilde{b} = \underset{\tilde{m}}{\min} \sum_{m=0}^{\tilde{m}} b_m$. \\
By Lemma \ref{lemma1} and compactness of the parameter space $\Psi$, we have 
\bean
&& \E\left[\sup_{\left|\bbeta_{\infty}\right| \leq \lambda^*} \left|\frac{\partial S_\psi(\bx; \eta)}{\partial \gamma} \right| \right] \leq 
\E\left[\sup_{\left|\bbeta_{\infty}\right| \leq \lambda^*} \left| C_1(\bx, \bbeta) \right| \right], \\ 
&& = \frac{2\sqrt{p}\lambda^* \sigma}{\eta \tilde{b}} \sqrt{2p\log\left(1+\frac{4}{\sqrt{p}\lambda^*}\right)} + \left(\frac{|\gamma| + \hat{b} }{\eta \tilde{b}} + \eta - \log \eta \right) = C_1 < \infty, \\
&& \E\left[\sup_{\left|\bbeta_{\infty}\right| \leq \lambda^*} \max_{j=1, \dots, p} \left|\frac{\partial S_\psi(\bx; \eta)}{\partial \bbeta_j} \right| \right] \leq \E\left[\sup_{\left|\bbeta_{\infty}\right| \leq \lambda^*} \left| C_2(\bx, \bbeta) \right| \right],
\\ 
&& =
\frac{7\lambda^*p^{3/2}\sigma^2}{\eta \tilde{b}}\sqrt{\log{\left(1+\frac{4}{\sqrt{p}\lambda^*}\right)}} + \left(\frac{|\gamma| + \hat{b}}{\eta \tilde{b}} + \eta - \log \eta \right)\sigma\sqrt{2 \log(2p)} = C_2 < \infty,  \\  
&& \E\left[\sup_{\left|\bbeta_{\infty}\right| \leq \lambda^*} \left|\frac{\partial S_\psi(\bx; \eta)}{\partial b_m} \right| \right] 
\leq \E\left[\sup_{\left|\bbeta_{\infty}\right| \leq \lambda^*} \left| C_3(\bx, \bbeta) \right| \right],
\\
&&= 1 - \log \eta + \frac{\hat{b}}{\tilde{b}^2\eta}\left(2\sqrt{p}\lambda^* \sigma
 \sqrt{2p\log\left(1+\frac{4}{\sqrt{p}\lambda^*}\right)} + |\gamma| + \hat{b}\right) = C_3 < \infty.
\eean 
By the convexity of the supremum and maximum,
\bean
\sup_{\left|\bbeta_{\infty}\right| \leq \lambda^*} \E \left [\left|\frac{\partial S_\psi(\bx; \eta)}{\partial \gamma} \right| \right] &\leq& \E\left[\sup_{\left|\bbeta_{\infty}\right| \leq \lambda^*} \left|\frac{\partial S_\psi(\bx; \eta)}{\partial \gamma} \right| \right]  \leq C_1,  \\ 
\sup_{\left|\bbeta_{\infty}\right| \leq \lambda^*} \max_{j=1, \dots, p} \E \left[\left|\frac{\partial S_\psi(\bx; \eta)}{\partial \bbeta_j} \right| \right] &\leq& \E\left[\sup_{\left|\bbeta_{\infty}\right| \leq \lambda^*} \max_{j=1, \dots, p} \left|\frac{\partial S_\psi(\bx; \eta)}{\partial \bbeta_j} \right| \right] \leq C_2, \\
\sup_{\left|\bbeta_{\infty}\right| \leq \lambda^*} \E \left[ \left|\frac{\partial S_\psi(\bx; \eta)}{\partial b_m} \right| \right] &\leq& \E\left[\sup_{\left|\bbeta_{\infty}\right| \leq \lambda^*} \left|\frac{\partial S_\psi(\bx; \eta)}{\partial b_m} \right| \right] \leq C_3.
\eean
Let $ C(\bx) = \max_k C_k(\bx, \bbeta)$, then $\E[C(\bx)] \leq \max_{k} C_k < \infty$. Therefore, $S_\psi$ is $C(\bx)$-Lipschitz in $\psi$, i.e., 
\bean
\|S_{\psi_1}(\bx; \eta) - S_{\psi_2}(\bx; \eta)\| \leq C(\bx) \|\psi_1 - \psi_2\|.
\eean
The function class of $S_{\psi}$ is denoted by $\mathcal{S} \in \{S_\psi 
 : \mathbb{R}^p \rightarrow \mathbb{R}\}$. By Lemma \ref{lemma:2}, we have 
\bean
N(\mathcal{S}, |\cdot |, \epsilon) \leq N(\Psi, \|\cdot\|, \epsilon/(2\mathbb{E}[C(\bx)])) < \infty.
\eean
By Lemma 3.1 of \cite{geer2000empirical}, $\mathcal{S}$ satisfies the ULLN (Uniform Law of Large Numbers), i.e.,
\bean 
\sup_{\psi \in \Psi} |\mathcal{R}_\eta^n(g_\theta, \bbeta) - \mathcal{R}_\eta(g_\theta, \bbeta)  | \overset{p}{\rightarrow}  0.
\eean
Thus, by the ULLN and assumption on $\mathcal{R}_\eta^n(g_{\hat{\theta}_n}, \hat{\bbeta}_n)$, we have
\bean
&& \mathcal{R}_\eta(g_{\hat{\theta}_n}, \hat{\bbeta}_n) - \mathcal{R}_\eta(g_{\theta_0}, \bbeta_0) \\ 
&\leq& \sup_{\psi \in \Psi} |\mathcal{R}_\eta^n(g_\theta, \bbeta) - \mathcal{R}_\eta(g_\theta, \bbeta)| + (\mathcal{R}_\eta^n(g_{\hat{\theta}_n}, \hat{\bbeta}_n) - \mathcal{R}_\eta^n(g_{\theta_0}, \bbeta_0)) + o_p(1) \\ 
&=& o_p(1) + o_p(1) + o_p(1) \overset{p}{\rightarrow} 0.
\eean
\end{proof}

\newpage 
\section{Optimal Portfolio Construction Algorithm with UPR}
\begin{algorithm}[!h] 
	\SetKwInOut{Input}{Input}
	\SetKwInOut{Output}{Output}
	\DontPrintSemicolon	
		\Input{$\bX = \{\bx_i\}_{i=1,\dots,n}$, $\mu_0$, ${\bf d}$, $\gamma_0$, and learning rate $\lambda$.}
		\Output{$\gamma, {\bf b}$, and $\bbeta$.}
Initialize with $t=0$,
\bean
\bbeta^{(0)} &=& (1/p, \dots, 1/p) \in \mathbb{R}^p, \\
\delta^{(0)}_m &\sim& U(0, 1), ~ m = 0, \dots, M \\ 
b_m^{(0)} &=& \delta^{(0)}_{m} - \delta^{(0)}_{m-1}, ~ m = 1, \dots, M,\\
\gamma^{(0)} &=& \gamma_0.
\eean
Let $L(\bbeta, \theta) = \mathcal{R}_\eta^n(g_\theta; \bbeta)$. For equality constraints of $\bbeta$, define 
$$
\mathcal{C}_1 = \{\bbeta ~ | ~ \hat{\bmu}^\top \bbeta = \mu_0\} , ~ \mathcal{C}_2 = \{\bbeta ~ | ~ {\bf 1}^\top\bbeta = 1\}
$$
\For{$t=0,1,\ldots$}{
For given $\lambda$, update $(\tilde{\bbeta}^{(t)}, \gamma^{(t)}, {\bf \delta}^{(t)})$ by the gradient descent based algorithm:
\bean 
\tilde{\bbeta}^{(t+1)} &=& \bbeta^{(t)} - \lambda \frac{\partial L}{\partial \bbeta} \\  
\gamma^{(t+1)} &=& \gamma^{(t)} - \lambda \frac{\partial L}{\partial \gamma} \\ 
{\bf \delta}^{(t+1)} &=& {\bf \delta}^{(t)} - \lambda \frac{\partial L}{\partial {\bf \delta}}.
\eean

Project $\bbeta$ into constraints. 
\[
\bbeta^{(t+1)} = \underset{\bbeta}{\mbox{argmin}}\|\tilde{\bbeta}^{(t+1)} - \bbeta\|^2 + I_{\mathcal{C}_1}(\bbeta) + I_{\mathcal{C}_2}(\bbeta),
\]
where $I_\mathcal{A}(x)$ returns 0 for $x \in \mathcal{A}$ and $\infty$, otherwise. \\

Let $(\eta_1$, $\eta_2)$ be Lagrange multipliers, updating rules are as follows:
\bean
\eta_1 &=& \frac{(\hat{\bmu}^\top\hat{\bmu}) ({\bf 1}^\top \tilde{\bbeta}^{(t+1)} - 1) -({\bf 1}^
\top\hat{\bmu})(\hat{\bmu}^\top \tilde{\bbeta}^{(t+1)} - \mu_0)}{p \hat{\bmu}^\top\hat{\bmu} - ({\bf 1}^
\top\hat{\bmu})^2} \\
\eta_2 &=& \frac{({\bf 1}^\top \hat{\bmu})({\bf 1}^\top {\tilde{\bbeta}^{(t+1)}}) - ({\bf 1}^\top \hat{\bmu}) - p(\hat{\bmu}^\top \tilde{\bbeta}^{(t+1)}) + p\mu_0}{({\bf 1}^
\top\hat{\bmu})^2 - p \hat{\bmu}^\top\hat{\bmu}} \\
\bbeta^{(t+1)} &=& \tilde{\bbeta}^{(t+1)} - \eta_1{\bf 1} - \eta_2 \hat{\bmu}.
\eean

For the positiveness of $\delta_m$, 
$\delta^{(t+1)}_m = \max(\delta_{m}^{(t+1)}, 0), ~ m=0, \dots, M.$
}
If the solution converges, $\gamma, \delta$, and $\bbeta$ are obtained. 
\caption{Optimal portfolio with UPR} \label{alg: optimal portfolio}
\end{algorithm}

\section{Numerical Studies}
\subsection{Simulation Study}
\begin{figure*}[t]
    \centering
    \subfigure[$\tau = 2/3 ~ (\lambda = 0.84)$]{\includegraphics[width=0.23\linewidth, height=3.8cm]{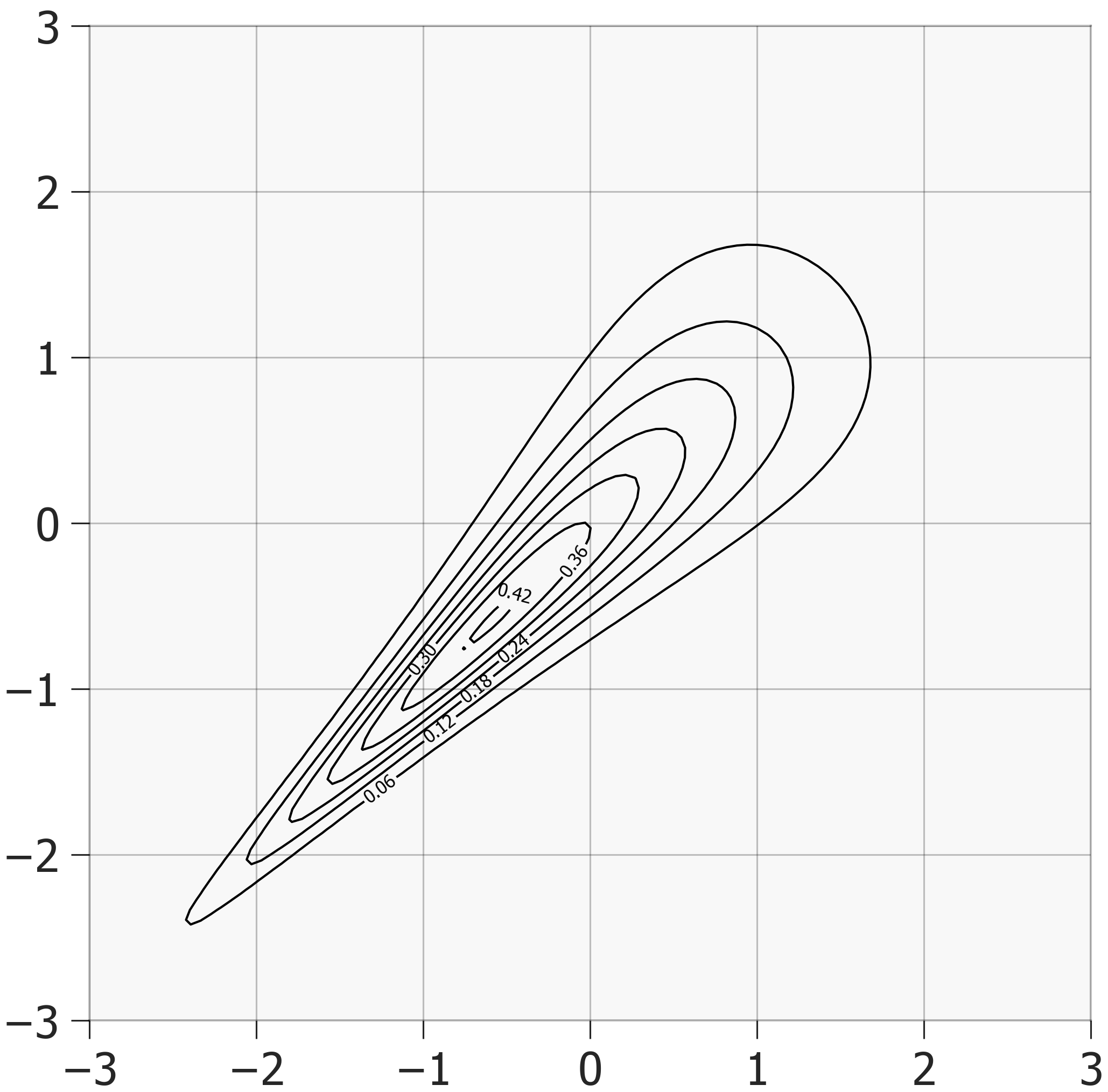}}   
    \subfigure[$\tau = 1/3 ~ (\lambda = 0.5)$]{\includegraphics[width=0.23\linewidth, height=3.8cm]{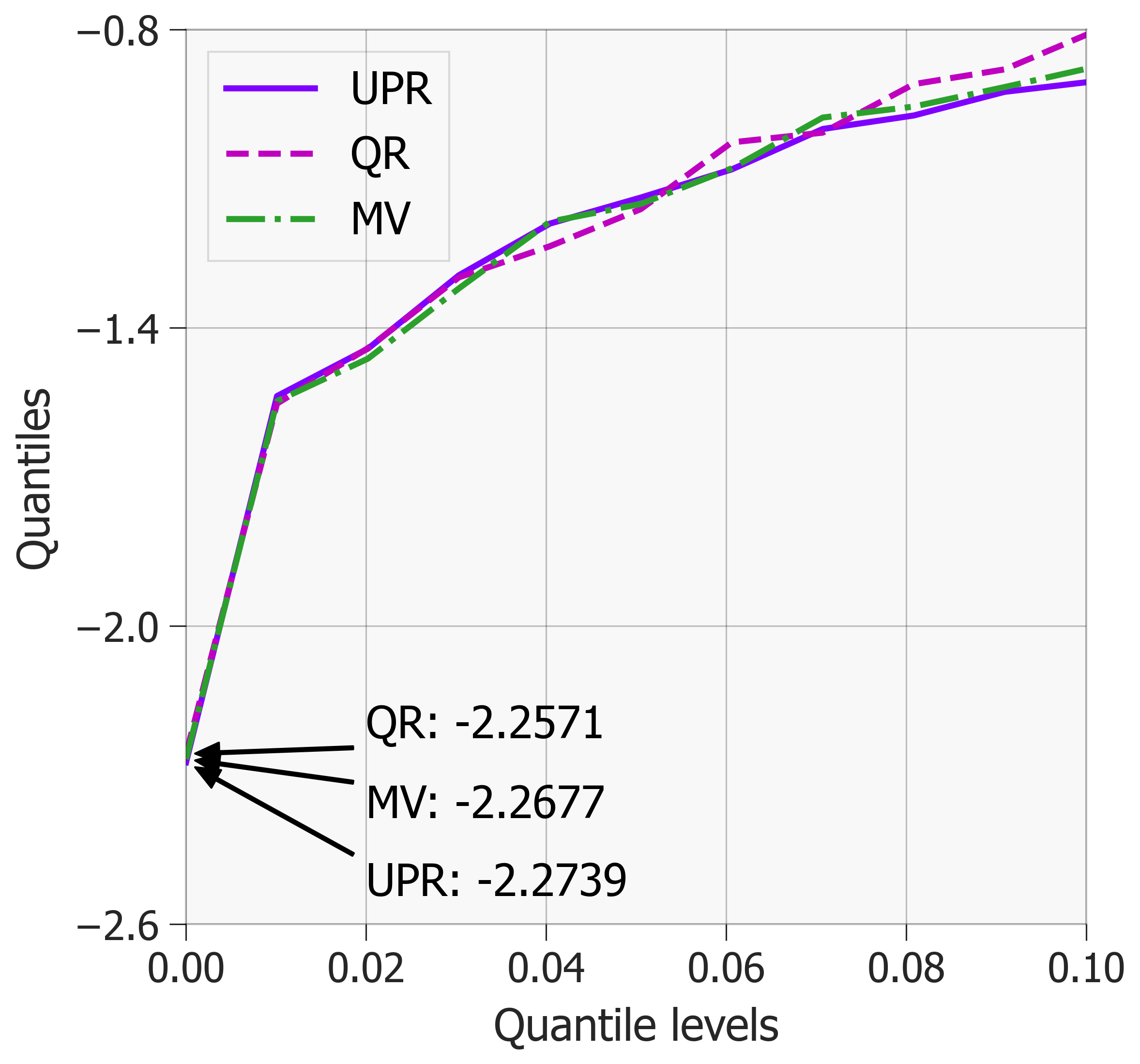}}
    \subfigure[$\tau = 2/3 ~ (\lambda = 0.84)$]{\includegraphics[width=0.23\linewidth, height=3.8cm]{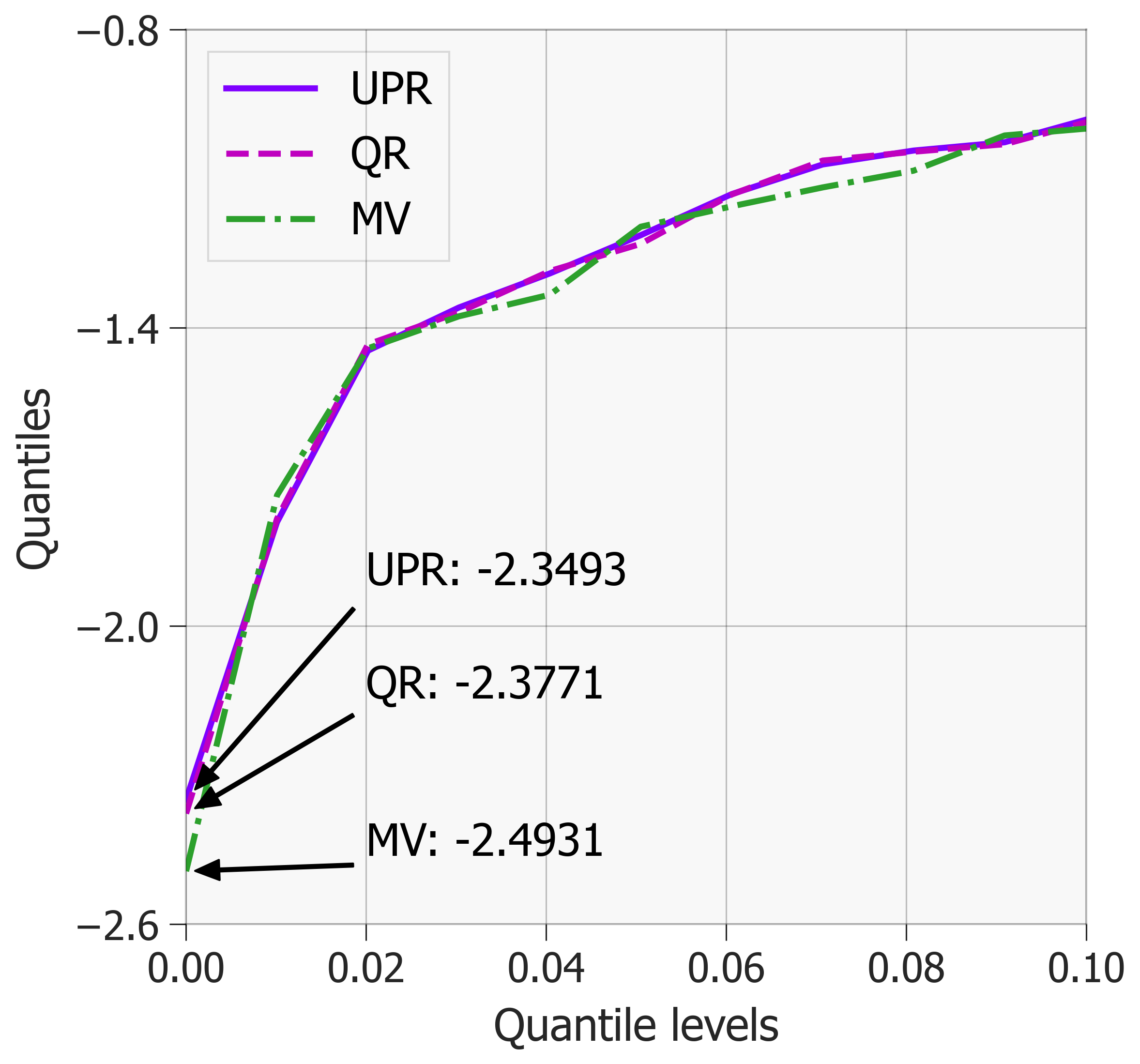}}
    \subfigure[$\tau=3/4 ~ (\lambda = 0.89)$]
    {\includegraphics[width=0.23\linewidth, height=3.8cm]{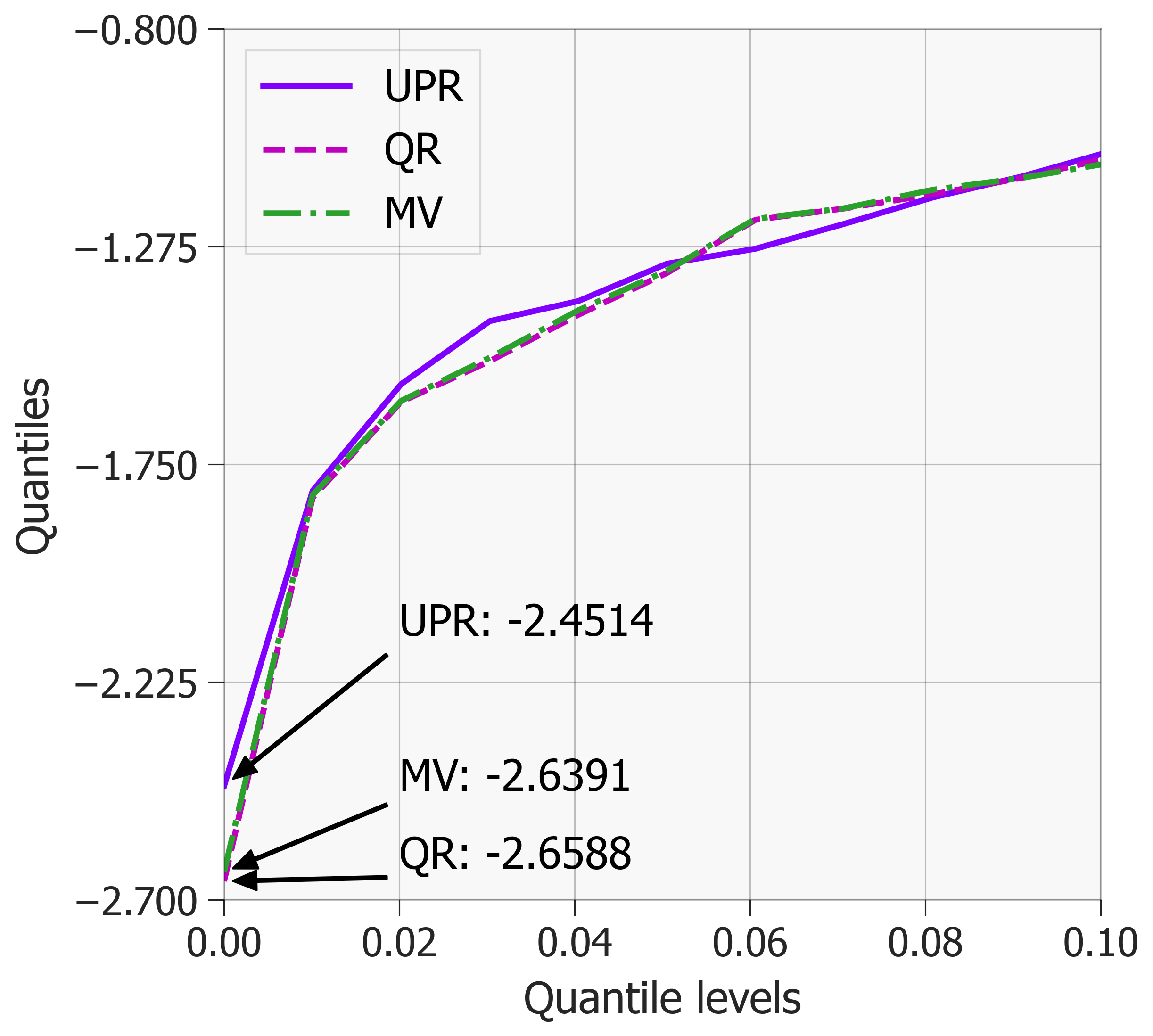}}
    \caption{(a) Joint PDF of the Clayton copula with $\tau=2/3$. (b) - (d)  Empirical quantile functions of portfolio returns with quantile level within the range $(0, 0.1]$.}
    \label{fig: simul_results}
\end{figure*}

Tail dependence serves as a comprehensive metric to gauge an extremal association between two distributions. In this simulation, we delve into the extreme risk arising from lower tail dependence. We evaluate our proposed portfolio model, UPR, against two benchmark portfolios: QR and MV.

We use three assets $X_1$, $X_2$, and $X_3$, where $X_1\sim N(0,1)$ and $X_2\sim N(0,1)$ are dependent, and $X_3\sim N(0,1.3^2)$ is independent of the two variables. We use the Clayton copulas with $\tau = 1/3, 2/3$ to model the lower tail dependence between $X_1$ and $X_2$. The coefficient of lower tail dependence is given by $\lambda = 2^{(\tau-1)/2\tau}$, where a higher $\tau$ value means a stronger lower tail dependence. Figure \ref{fig: simul_results} (a) shows the contour plots of the joint PDF of $X_1$ and $X_2$ at $\tau=2/3$. We fit each portfolio model with 300 random samples from the above setting with an equal expected return.

Figure \ref{fig: simul_results} (b) and (c) show the empirical quantile functions of portfolio returns at low quantile levels. At $\tau = 1/3$, UPR incurs the maximum loss of $2.274$, compared to QR's $2.257$ and MV's $2.268$. Conversely, at $\tau = 2/3$, UPR records its maximum loss of $2.349$, compared to QR's $2.377$ and MV's $2.493$. As the tail dependency strengthens and extreme loss increases, the UPR portfolio prioritizes minimizing the maximum loss with $\phi$ over $\nu_{0.1}$ and $\sigma$-risk, to minimize UPR. Furthermore, we examine the out-of-sample performance as increasing $\tau$s. We construct each portfolio model at $\tau=2/3$ and then measure the out-of-sample performance using 300 random samples at $\tau=3/4$.  Figure \ref{fig: simul_results} (d) illustrates the empirical quantile function of portfolio returns in the out-of-sample scenario. Regarding extreme risk in the out-of-sample, the UPR portfolio surpasses other models, observing a maximum loss of 2.451, which is lower compared to 2.659 for QR and 2.639 for MV.

\subsection{Benchmark Models}
We compare the out-of-sample performances of our proposed model (UPR) with seven comparable models \footnote{The optimal weights of comparable models are computed using \textsf{quantreg} \citep{koenker2018package} and \textsf{lpSolve} packages with \textsf{R}.} as follows: 
\ben
    \item \textbf{QR} 
    \cite{bassett2004pessimistic}: the quantile Regression model with $\alpha=0.1$ in \eqref{eq: qr_model}. 
    \item \textbf{EW} \cite{demiguel2009optimal}: an equally weighted portfolio that invests in equal proportions in all assets.
    \item \textbf{CQR1} \cite{bassett2004pessimistic}: the composite Quantile Regression model that estimates three quantile levels $\{0.1, 0.5, 0.9\}$  with weight vector $(1/3, 1/3, 1/3)$  in \eqref{obj: weighted alpha portfolio}.
    \item \textbf{CQR2} \cite{bassett2004pessimistic}: the composite Quantile Regression model that estimates three quantile levels $\{0.01, 0.1, 0.5, 0.9\}$ with weight vector $(0.4, 0.3, 0.2, 0.1)$ in \eqref{obj: weighted alpha portfolio}.
    \item \textbf{EO} \cite{chen2022robust}: the portfolio with CVaR based on the linear programming.
    \item \textbf{RO} \cite{chen2022robust}: the distributional robust optimization version of `EO'.
    \item \textbf{MV} \cite{markowitz1952portfolio}: the mean-variance portfolio.  
\een

\subsection{Performance Measures} 
\bed 
    \item \textbf{CW}: Cumulative Wealth calcuated with initial return $r_0 = 1$, i.e., $\mbox{CW} = \sum_{t=0}^T r_t$, where $r_t$ is the $t$-th portfolio return and $T$ is the total evaluation length.
    \item \textbf{MDD}: Maximum DrawDown calculated as the ratio of the most observed loss from the previous peak to the trough to its peak.
    \item \textbf{MaxLoss}: Maximum Loss, i.e., $-1 \times \min(r_t)$.
    \item \textbf{CVaR$_{0.1}$}: CVaR at level $0.1$ calculated as the negative average of returns lower than $0.1$-quantile.
    \item \textbf{SR}: Sharpe Ratio calculated as the ratio of the average of returns to their standard deviation $\sum_{t=1}^T r_t / (T\times\text{sd}(r_t))$. 
\eed

\subsection{Statistical Significance Test to Compare Two Sharpe Ratios}
We provide the statistical difference in portfolio performance between our proposed method and benchmarks in Table \ref{tab:result_rda}. To test whether the performances of two portfolios are distinguishable, we compared two Sharp ratios of portfolios \cite{demiguel2009optimal, memmel2003performance} where the null hypothesis of test $H_0: SR_i = SR_j$. The test statistics can be calculated as follows: 
\begin{align}
\hat{Z}_{ij} &= \frac{\hat{\mu}_i \hat{\sigma}_j - \hat{\mu}_j \hat{\sigma}_i}{\sqrt{\vartheta}} \sim N(0, 1) \nonumber \\ 
\vartheta &= \frac{1}{L}\left(2 \hat{\sigma}_i^2\hat{\sigma}_j^2 - 2\hat{\sigma}_i\hat{\sigma}_j\hat{\sigma}_{i,j} + \frac{1}{2}\left(\hat{\mu}_i^2\hat{\sigma}_j^2 + \hat{\mu}_j^2\hat{\sigma}_i^2\right) - \frac{\hat{\mu}_i\hat{\mu}_j}{\hat{\sigma}_i\hat{\sigma}_j}\hat{\sigma}_{i,j}^2 \right), \nonumber
\end{align}
where $\hat{\mu}_k$, $\hat{\sigma}_k$, and $\hat{\sigma}_{k,l}$ denote average return, standard deviation of return of $k$-th portfolio and covariance between $k$th and $l$th portfolio returns. $L$ is the out-of-sample size of the portfolio return. 
Through test results, our model significantly outperforms the others in KOSPI200 and S\&P500 (excluding EW). In CSI500, there is no significant difference among models in SR. 

\newpage
\subsection{Quantile Discrepancy in Out-of-sample}
\begin{figure}[h]
    \centering
    \subfigure[UPR]{\includegraphics[width=0.9\linewidth]{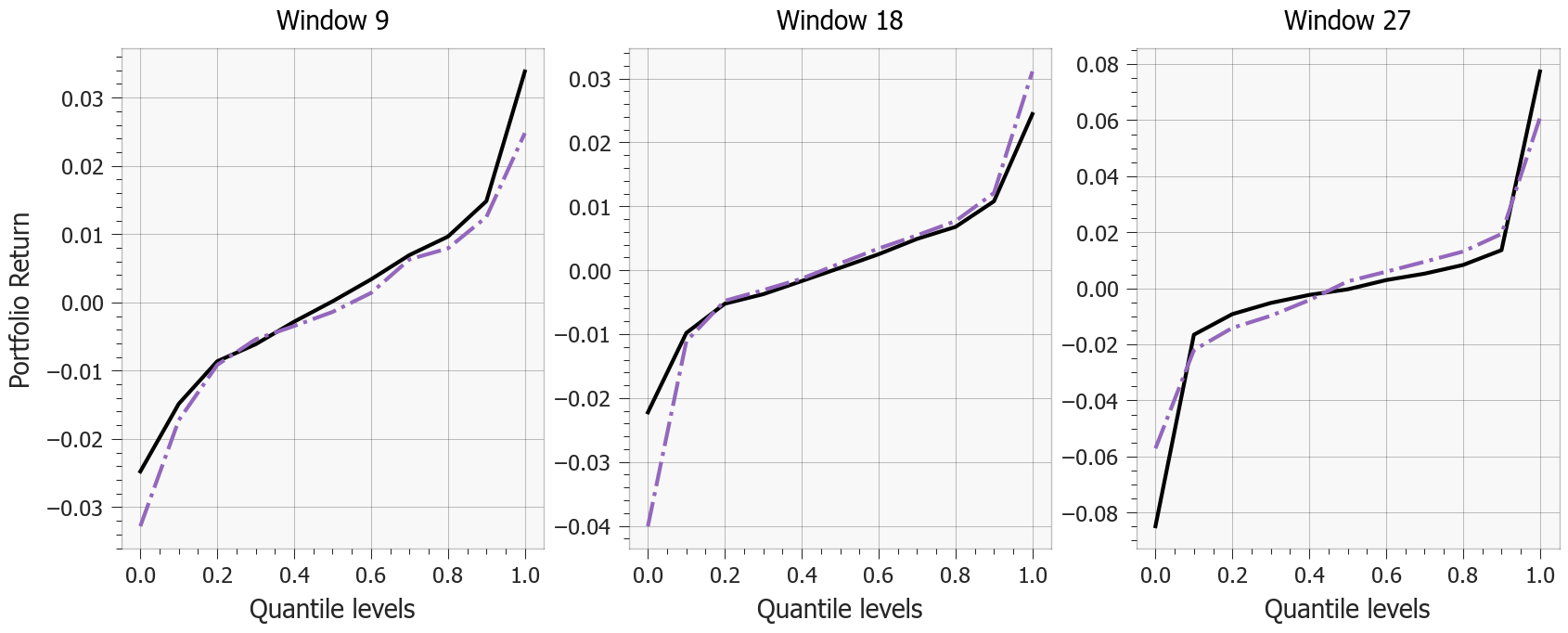}}
    \subfigure[CQR1]{\includegraphics[width=0.9\linewidth]{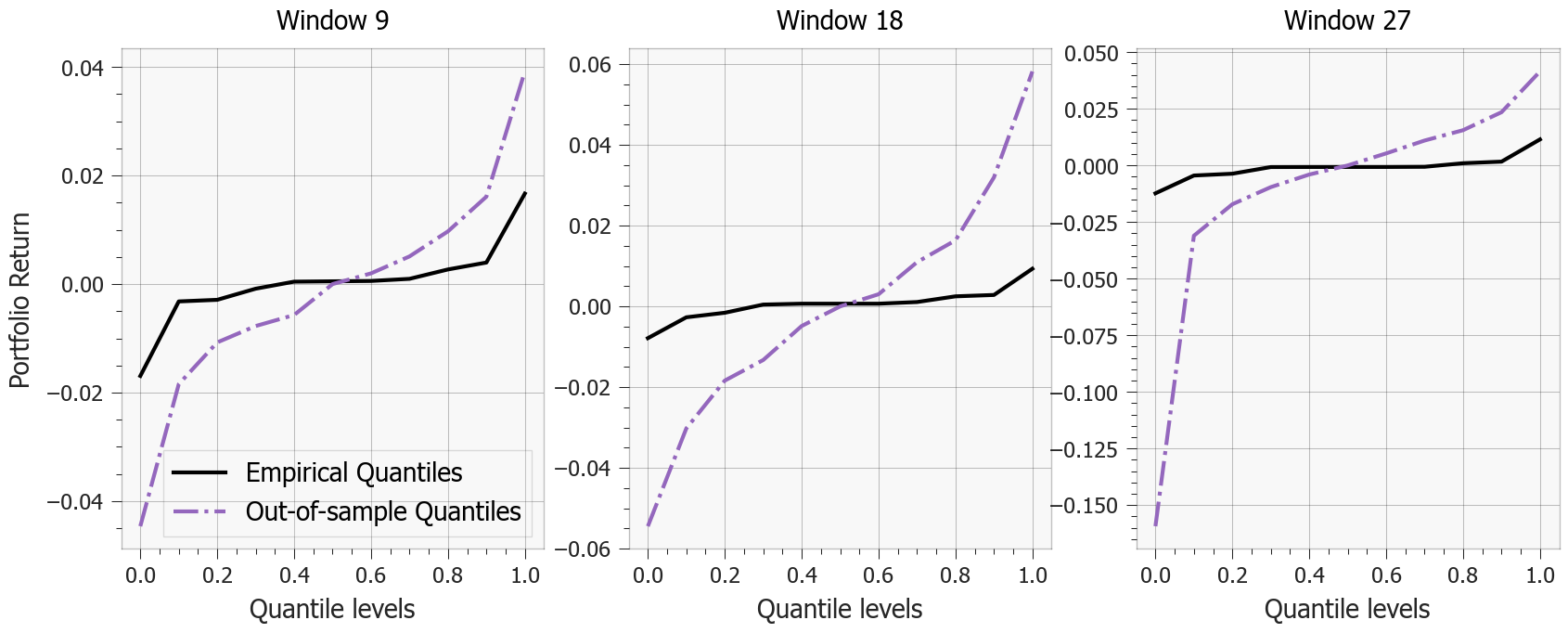}}
    \subfigure[CQR2]{\includegraphics[width=0.9\linewidth]{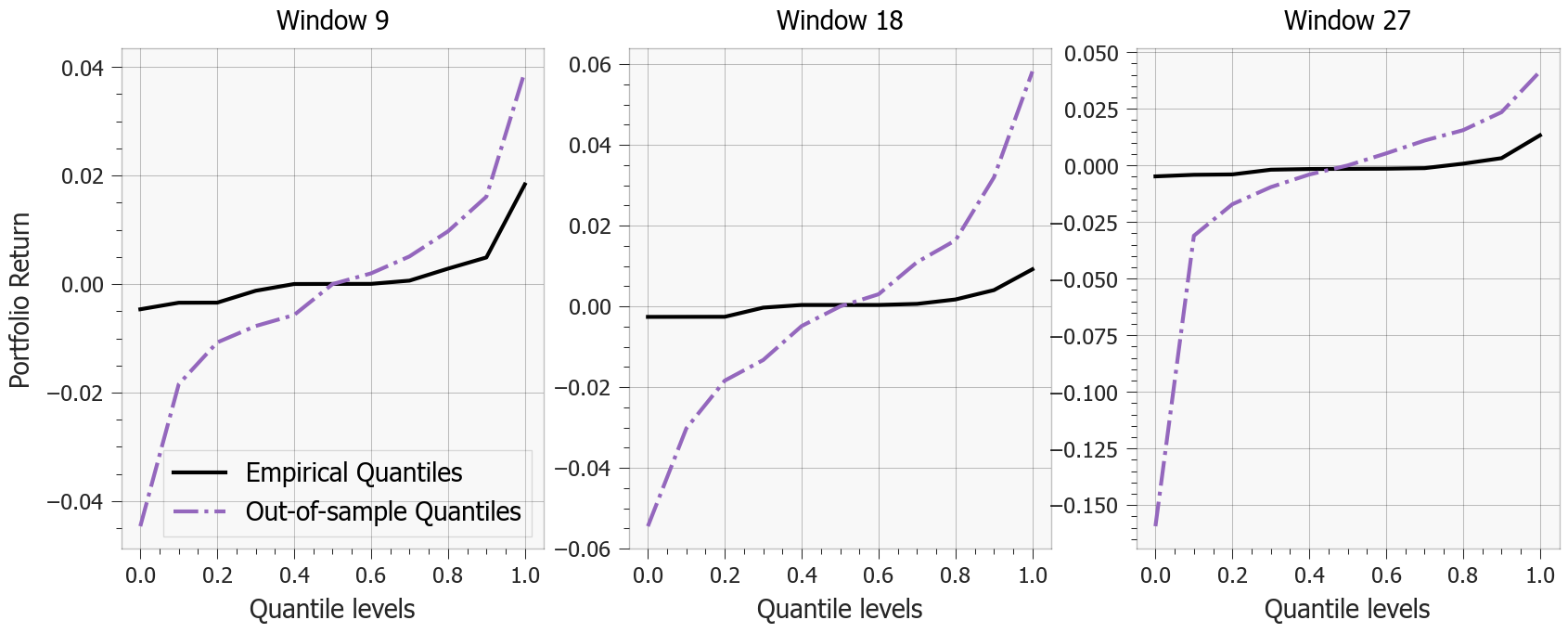}}
    \caption{Out-of-sample evaluation via quantile function matching in KOSPI200 dataset.}
    \label{fig:qf}
\end{figure}

Figure \ref{fig:qf} presents the results of quantile matching of the UPR portfolio in Section \ref{sec: num}. The lower tail estimation outperforms the upper tail (see blue dashed and black solid lines) because $\phi(t)$ in Proposition \ref{prop: fubini} gives larger weights to lower quantiles than higher quantiles. Consequently, this weighting approach contributes to the UPR portfolio's enhanced efficacy in mitigating extreme losses. Providing the quantile discrepancy between in-sample and out-of-sample, the entire quantile estimation, which reflects the most conservative return distribution and our DRO property, is more adequate for robust portfolio optimization.

\end{document}